\documentclass[aps,prd,superscriptaddress,twocolumn,nofootinbib,floatfix]{revtex4-2}
\usepackage[normalem]{ulem}
\usepackage{natbib}
\usepackage{graphicx}
\usepackage{amsfonts}
\usepackage{amsmath,amssymb} % \colleneqq for definition symbol
\usepackage{bm}
\usepackage{xcolor}
\usepackage{hyperref}
\usepackage[caption=false]{subfig}
\usepackage{tabularx}
\usepackage{mathtools}
\usepackage{xfrac} % for slanted fractions
\usepackage{orcidlink}
\usepackage{makecell} % For multiline cells in table

\renewcommand{\sp}{\epsilon} %%Small Parameter = mass ratio
\newcommand{\mr}{\epsilon} %% Mass Ratio
\newcommand{\mycomment}[1]{} %Command to effectively comment out blocks of code in TeX

\newcommand{\avg}[1]{\left\langle #1 \right\rangle}

\newcommand{\osc}[1]{\Breve{#1}}
\newcommand{\nit}[1]{\Tilde{#1}}
\newcommand{\PD}[2]{\frac{\partial #1}{\partial #2}}
\newcommand{\HOT}[1]{\mathcal{O}( \sp ^{ #1 } )} 

\definecolor{aldo}{RGB}{210, 190, 100}

\renewcommand{\vec}[1]{\mathbf{#1}}

\begin{document}

%\preprint{}

\title{Efficient Eccentric Effective-One-Body Dynamics via Near-Identity Averaging Transformations}


\newcommand{\COG}{Center of Gravity, Niels Bohr Institute, Blegdamsvej 17, 2100 Copenhagen, Denmark}
\newcommand{\AEI}{Max Planck Institute for Gravitational Physics (Albert Einstein Institute), Am M\"uhlenberg 1, Potsdam 14476, Germany}
\newcommand{\UofM}{Department of Physics, University of Maryland, College Park, MD 20742, USA}

\author{Philip Lynch \orcidlink{https://orcid.org/0000-0003-4070-7150}}
\email[Email: ]{philip.lynch@aei.mpg.de}
\affiliation{\AEI}
%\thanks{}

\author{Alessandra Buonanno \orcidlink{https://orcid.org/0000-0002-5433-1409}}
\affiliation{\AEI}
\affiliation{\UofM}
\author{Aldo Gamboa \orcidlink{https://orcid.org/0000-0001-8391-5596}} 
\affiliation{\AEI}
\author{Maarten van de Meent \orcidlink{https://orcid.org/0000-0002-0242-2464}}
\affiliation{\COG}
\affiliation{\AEI}

\date{\today}

\begin{abstract}

Next-generation gravitational-wave detectors, such as LISA, the Einstein Telescope, and Cosmic Explorer, will require accurate and efficient models of long-lived black hole binary signals, including those with significant eccentricity. A challenge for eccentric effective-one-body models is the cost of resolving rapidly oscillating orbital dynamics over the inspiral, particularly for low-mass and large-mass-ratio systems. We address this by recasting the nonspinning eccentric effective-one-body equations of motion in terms of osculating orbital elements and then applying near-identity averaging transformations to eliminate the fast orbital-timescale structure during the inspiral. Each order in this procedure suppresses the oscillatory behavior by one factor of the ratio of orbital to radiation-reaction timescales. The resulting averaged dynamics are evolved on the radiation-reaction timescale, and then the system is mapped back to the full EOB dynamics for the final transition to plunge. This reduces the cost of the inspiral dynamics by up to two orders of magnitude, eliminating this as the primary bottleneck for long waveforms. The overall waveform-generation speedup spans $1.5 - 8 \times$, motivating the development of more efficient waveform-generation methods. We also validate the accuracy of the method by comparing waveforms generated from the averaged and full effective-one-body dynamics across a broad region of parameter space for moderate to large eccentricities. Carrying out this averaging procedure to next-to-next-to-leading order is needed to accurately model comparable-mass binaries, yielding mismatches $\leq 8.05 \times 10^{-5}$. These results establish near-identity averaging as a practical route to efficient eccentric effective-one-body inspirals, and provide a foundation for further extensions to low-eccentricity and spinning waveform models.
\end{abstract}

\keywords{Gravitational Waves, Black hole binaries, Effective-One-Body}

\maketitle


\section{Introduction \label{section:Intro}}
Since the first detection of gravitational waves from a coalescing binary black hole (BBH)\cite{LIGOScientific:2016aoc}, the LIGO-Virgo-KAGRA collaboration has confirmed over 390 gravitational wave (GW) signals from merging compact objects such as stellar mass black holes and neutron stars \cite{LIGOScientific:2018mvr,LIGOScientific:2020ibl,KAGRA:2021vkt,LIGOScientific:2025slb,LIGOScientific:2026wfs}. Waveform modeling has become a central pillar of GW astronomy: accurate waveform models are needed not only for detection, but also for precise inference of source properties and for tests of general relativity. Detector sensitivity will continue to improve with future ground-based detectors like the Einstein Telescope \cite{ET:2025xjr} and Cosmic Explorer \cite{Reitze:2019iox} as well as space-based detectors such as the Laser Interferometer Space Antenna (LISA) \cite{LISA:2024hlh,LISAConsortiumWaveformWorkingGroup:2023arg}, TianQin \cite{TianQin}, and Taiji \cite{Taiji}. As such, previously subdominant waveform systematics will become increasingly important \cite{Nagar:2023zxh,Dhani:2024jja}. This is particularly true for long-lived signals which accumulate many inspiral cycles in band.

Including orbital eccentricity in waveform models has become increasingly important as evidence for eccentricity in detected signals has mounted, though not to the point of a definitive detection of eccentricity \cite{Gayathri:2020coq,Romero-Shaw:2021ual,Romero-Shaw:2022xko,Iglesias:2022xfc,Ramos-Buades:2023yhy,Gupte:2024jfe,Morras:2025xfu,Planas:2025jny,Xu:2025ajj,Gupte:2026whi,Zeeshan:2026pga}. While gravitational radiation tends to circularize binaries during their inspiral \cite{Peters:1963ux,Peters:1964zz,Hinder:2007qu,Sperhake:2007gu}, a range of astrophysical formation channels in globular clusters and active galactic nuclei can produce systems that retain measurable eccentricity when they enter the observational band of current and future detectors \cite{Downing_2010,Antonini:2012ad,OLeary:2008myb,Petrovich:2017otm,Gondan:2017wzd,Takatsy:2018euo}. This is especially relevant for next-generation observatories which will be more sensitive to the inspiral where eccentricity effects are most pronounced \cite{LISAConsortiumWaveformWorkingGroup:2023arg}.

 As such, many waveform modeling frameworks have been extended to include orbital eccentricity \cite{Huerta:2016rwp,Huerta:2017kez,Hinder:2017sxy,Chattaraj:2022tay,Manna:2024ycx,Paul:2024ujx,Islam:2022laz,Nee:2025nmh,Maurya:2025shc,Morras:2025nlp,Planas:2025feq,Ramos-Buades:2026kbq,Morras:2026fho,Islam:2026blk}. In this work, we focus on the effective-one-body (EOB) framework \cite{Buonanno:1998gg,Buonanno:2000ef,Damour:2000we,Buonanno:2005xu,Barausse:2009xi}. It provides a powerful description of compact-binary dynamics by resumming information from post-Newtonian (PN) \cite{Nagar:2011fx,Bini:2012ji,Balmelli:2013zna,Damour:2015isa,Khalil:2020mmr,Khalil:2021txt,Gamboa:2024imd}, gravitational self-force (GSF) \cite{Antonelli:2019fmq,vandeMeent:2023ols,Leather:2025nhu}, and recently post-Minkowskian (PM) theory \cite{Damour:2016gwp,Antonelli:2019ytb,Khalil:2022ylj,Buonanno:2024byg,Buonanno:2024vkx,Damour:2025uka}, into an effective system of a test mass moving in a deformed black hole spacetime. EOB models are then calibrated against numerical relativity (NR) simulations to ensure agreement in the late inspiral and merger-ringdown \cite{Pan:2009wj,Buonanno:2009qa,Bohe:2016gbl}. The EOB paradigm has proven highly accurate and forms the basis of several state-of-the-art waveform models used in GW data analysis \cite{Nagar:2018zoe,Nagar:2019wds,Gamba:2021ydi,Taracchini:2012ig,Taracchini:2013rva,Purrer:2015tud,Babak:2016tgq,Bohe:2016gbl,Cotesta:2018fcv,Pompili:2023tna,Khalil:2023kep,Ramos-Buades:2023ehm,Mihaylov:2023bkc} with recent extensions including orbital eccentricity \cite{Chiaramello:2020ehz,Nagar:2020xsk,Nagar:2021gss,Nagar:2021xnh,Placidi:2021rkh,Albanesi:2021rby,Albanesi:2022ywx,Albanesi:2022xge,Albanesi:2023bgi,Placidi:2023ofj,Nagar:2024dzj,Nagar:2024oyk,Cao:2017ndf,Liu:2019jpg,Liu:2021pkr,Ramos-Buades:2021adz,Gamboa:2024hli,Gamboa:2026jht,Pompili:2026yxq} and spin precession \cite{Liu:2023ldr,Gamba:2024cvy,Albanesi:2025txj,Gamboa:2026ybr}. 

This accuracy comes at a cost: the inspiral dynamics must be obtained by numerically solving the EOB Hamiltonian equations of motion.
In the quasicircular case, this can be bypassed with the ``postadiabatic approximation'' \cite{Damour:2007yf,Damour:2012ky,Nagar:2018gnk}: one perturbs around a circular orbit in order to recast the EOB equations of motion into a set of algebraic equations that can be solved analytically \cite{Rettegno:2019tzh,Riemenschneider:2021ppj,Mihaylov:2021bpf}. This approach is ill suited to eccentric systems, so one must integrate the inspiral dynamics numerically. This is particularly costly for low-mass, large-mass-ratio systems whose inspirals contribute many waveform cycles within the detector band.

To address this problem, we adapt techniques that have proven highly successful in the modeling of eccentric extreme-mass-ratio inspirals (EMRIs) \cite{Chua:2020stf,Katz:2021yft,Speri:2023jte,Chapman-Bird:2025xtd}. We first recast the eccentric EOB dynamics in terms of osculating orbital elements (OOE) \cite{Pound:2007th,Gair:2010iv} (also known as the method of variation of constants \cite{taff1985celestial,beutler2005volume1}) using a quasi-Keplerian parametrization à la Ref.~\cite{Hinderer:2017jcs}. This makes explicit the separation between the slowly evolving orbital elements and the rapidly varying orbital phase. We then apply near-identity (averaging) transformations (NITs) \cite{Kevorkian1987} to recast the short-timescale oscillatory structure from the inspiral dynamics as higher-order secular information needed for faithful waveform generation \cite{vandeMeent:2018cgn,McCart:2021upc,Lynch:2021ogr,Pound2022,Lynch:2022zov,Lynch:2023gpu,Drummond:2023wqc,Lynch:2024ohd,Drummond:2026haw}. The result is a new set of averaged equations of motion that can be solved on the radiation-reaction timescale rather than the orbital timescale. 
Unlike in the EMRI case, we do not expand in the mass ratio, but rather in the small parameter $\epsilon$ that characterizes the separation of timescales between the orbital and radiation-reaction timescales. This is suppressed not only by $\nu$ but also by the fact that radiation reaction begins at 2.5PN order \cite{Peters:1963ux,Peters:1964zz}. 
We also extend the calculation beyond the next-to-leading order terms needed for EMRIs \cite{Hinderer:2008dm, Burke:2023lno}, to include next-to-next-to-leading order terms.
The resulting averaged dynamics are then mapped back to the full EOB dynamics for the final transition to plunge.

We validate the method by testing it on an inspiral-only, nonspinning EOB model. We compare waveforms generated from the averaged dynamics against those from the full EOB evolution across a range of eccentric configurations. We find that next-to-next-to-leading order is needed to model comparable-mass systems accurately with the resulting model reproducing the full dynamics to mismatches $\leq 8.05 \times 10^{-5}$ while substantially reducing computational cost by a factor of $1.5 - 8 \times$. NITs thus provide a practical route to accelerating eccentric EOB waveform generation without sacrificing the accuracy required for data-analysis applications.

The remainder of this paper is organized as follows. In Sec.~\ref{section:EOB_Dynamics}, we review the nonspinning eccentric EOB inspiral dynamics used in this work. In Sec.~\ref{section:OsculatingOrbitalElements}, we rewrite these dynamics using osculating orbital elements. In Sec.~\ref{section:NITs}, we derive the near-identity averaged equations of motion through second postadiabatic order. In Sec.~\ref{section:Waveforms}, we describe waveform generation from both the full and averaged dynamics, and in Sec.~\ref{section:Implementation} we summarize the numerical implementation. We present accuracy and runtime comparisons in Sec.~\ref{section:Results}, and conclude in Sec.~\ref{section:Discussion} with a discussion of limitations and future extensions. In this work, we use geometric units where $G = c = 1$.

\section{Nonspinning effective-one-body dynamics} \label{section:EOB_Dynamics}
We now describe a two-body system consisting of a  primary of mass $m_1$ and a secondary of mass $m_2$. For convenience, we define the total mass $M = m_1 + m_2$ and the reduced mass $\mu = m_1 m_2 / M$.
In the EOB framework, we model the binary as a test mass moving in a deformed black hole spacetime where the deformation parameter is the symmetric mass ratio $\nu =  \mu / M = m_1 m_2 / (m_1 + m_2)^2$. We focus on the nonspinning sector, but our methods extend straightforwardly to (anti-)aligned spin effects.

The metric of this deformed spacetime in EOB coordinates $(t, r,\phi)$ is given by:
\begin{equation}
	ds^2 = -A(\nu;r) dt^2 + \frac{dr^2}{A(\nu;r) \bar{D}(\nu;r)} + r^2 d\phi^2
\end{equation}
Note that in the test mass limit $\nu \rightarrow 0$, $A \rightarrow 1 - 2 M/r$ and $\bar{D} \rightarrow 1$, recovering the Schwarzschild metric. Many choices for $A$ and $\bar{D}$ exist in the literature, but here we adopt those of \texttt{SEOBNRv5} \cite{Pompili:2023tna,Khalil:2023kep}.

Note that the test mass does not follow a geodesic in the deformed spacetime as it is slowly inspiraling due to radiation reaction. Instead, the test mass follows a perturbed orbit satisfying:
\begin{equation}
u^\alpha p_\alpha + Q(\nu;r,p_{r_*}) = g^{\beta \alpha} p_\beta p_\alpha + Q(\nu;r,p_{r_*}) =  -1.
\end{equation}
Here, $u^\alpha$ and $p_\alpha$ represent the four-velocity and four-momentum, respectively,  of the test mass (all normalized by a factor of $\mu$), $p_{r_*} =  \xi(r) p_r $ is the tortoise radial momentum, where $\xi(r) = A \sqrt{\bar{D}}$.
This $Q$ potential captures the effect of the slow inspiral on the conservative dynamics.
 As such, $Q \rightarrow 0$ in the test mass limit. It is given as a PN series that appears at 3PN order and is expressed as an expansion in $p_{r_*}$ (or $p_r$) and has the form
 \begin{equation} \label{eq:Q_Pot}
	Q = Q_4(\nu,r) p_{r_*}^4 + Q_6(\nu,r) p_{r_*}^6 + Q_8(\nu,r) p_{r_*}^8 + \mathcal{O}(p_{r_*}^{10}).
\end{equation}
 Again, we use the same expressions as \texttt{SEOBNRv5} where the full expressions for $Q_4,Q_6,$ and $Q_8$ for nonspinning binaries can be found in Appendix A of Ref.~\cite{Pompili:2023tna}.

 From the above equation, one derives the effective Hamiltonian by noting that $p_t = - H_{\text{eff}}$, converting from $p_r$ to $p_{r_*}$ and rearranging:
\begin{equation} \label{eq:Heff}
 H_{\text{eff}}^2 = p_{r_*}^2 + A(\nu,r)\left[1 + \frac{p_\phi^{2}}{r^2} + Q(\nu;r,p_{r_*})\right]
\end{equation}

 The effective Hamiltonian is related to the real EOB Hamiltonian by the EOB energy map:
\begin{equation}
	H_\text{EOB} = M \sqrt{1 + 2 \nu (H_\text{eff} -1)}.
\end{equation}
From this Hamiltonian, one obtains the equations of motion via Hamilton's equations:
\begin{align} \label{eq:EOB_EOM}
	\begin{split}
	\frac{d r}{dt} &= \xi(r) \frac{\partial H_{\text{EOB} }}{\partial p_{r_*}}, \quad \frac{d \phi}{dt} = \frac{\partial H_{\text{EOB}}}{\partial p_\phi} = \Omega_\phi,
	\end{split}\\
\begin{split}
	\frac{d p_{r_*}}{dt} &= - \xi(r) \frac{\partial H_{\text{EOB}}}{\partial r} + \xi(r) F_r, \quad \frac{d p_\phi}{dt} = F_\phi,
	\end{split}
\end{align}
where $F_r$ and $F_\phi$ are the radiation-reaction force components.
We use the nonspinning limit of the 3PN radiation-reaction force components from Ref.~\cite{Gamboa:2024imd} expressed in EOB coordinates $(r,p_r,p_\phi)$; we provide them in Appendix~\ref{section:RRForce}.

\section{Osculating Orbital Elements} \label{section:OsculatingOrbitalElements}
\subsection{Recasting into quasi-Keplerian parametrization}\label{section:QuasiKeplerianParametrization}
We now recast these equations into a set of osculating orbital element equations using the following quasi-Keplerian parametrization:
\begin{equation} \label{eq:r_of_p_e_xi}
	r =  \frac{p M }{1 + e \cos \zeta},
\end{equation}
where $p$ is the semilatus rectum, $e$ is the eccentricity, and $\zeta$ is the relativistic anomaly. This is commonly decomposed into $\zeta =\chi - w$, where $\chi$ is a radial phase parameter and $w$ is the argument of periapsis.
The semilatus rectum and eccentricity can be related to the radial roots, $ r_1 $ and $ r_2 $, via:
\begin{equation}
	p = \frac{2 r_1 r_2}{M(r_1 + r_2)}, \quad 
	e = \frac{r_1 - r_2}{r_1 + r_2}.
\end{equation}
These can be found by taking the roots of the radial potential, i.e.:
\begin{equation}
\frac{d r}{dt} = \xi(r) \frac{\partial H_\text{EOB}}{\partial p_{r_*}} \Bigg \vert _{r_1,r_2} = 0.
\end{equation}

In this parametrization, one can find expressions for the other constants of motion, such as the effective energy and angular momentum \cite{Hinderer:2017jcs}:
\begin{subequations}
	\begin{equation}
		H_{\text{eff}} =-p_t =  \frac{2 \sqrt{e} \sqrt{A(\nu;r_1) A(\nu;r_2)}}{\sqrt{(1+e)^2 A(\nu;r_2) - (1 - e)^2 A(\nu;r_1)}},
	\end{equation}
	\begin{equation} \label{eq:pphi_of_p_e}
		p_\phi = \sqrt{\frac{p^2 M^2 (A(\nu;r_2) - A(\nu;r_1))}{(1-e)^2 A(\nu;r_1) - (1+e)^2 A(\nu;r_2)}}.
	\end{equation}
\end{subequations}
Inverting the original expression for $H_{\text{eff}}(\nu;r,p_{r_*},p_\phi)$ yields an expression for $p_{r_*}(\nu;p,e,\zeta)$. 
This can be done exactly in the test mass limit or when the Q potential consists only of its leading order term \cite{Hinderer:2017jcs}. However, including more terms in the Q potential generally precludes an exact analytical inverse, so one must numerically root find for $p_{r_*}(\nu;p,e,\zeta)$. We avoid this by exploiting that $Q$ is a power series expansion in $p_{r_*}$ and perturbatively invert the series to obtain an approximate expression for $p_{r_*}(\nu;p,e,\zeta)$. 
Since Q is known to $\mathcal{O}(p_{r_*}^8)$, this approximation remains very accurate even at large eccentricities. 
We give an explicit expression for this expansion in Appendix~\ref{section:prCoefficients}.

\subsection{The method of osculating orbital elements}

To uniquely specify an eccentric, planar orbit, one need specify $p$, $e$, and $w$. We denote these as the orbital elements $I^A = (p,e,w)$.  In the absence of radiation reaction, they are constant.

We now  recast the EOB equations of motion into evolution equations for the orbital elements via the method of osculating orbital elements (OOE) \cite{Pound:2007th,Gair:2010iv}. At each instant in time ($t$) one maps the worldline of the inspiraling test mass $x^\alpha$ onto a tangent conservative orbit sharing the same orbital elements $I^A$ via the osculating conditions:
\begin{subequations}
	\begin{equation}
		x^\alpha(t) = x_\text{cons}^\alpha(I^A(t), t)
	\end{equation}
	
		\begin{equation}
		\frac{d x^\alpha}{d t} (t) = \PD{x_\text{cons}^\alpha}{t}(I^A(t), t).
	\end{equation}
\end{subequations}
From these, one can derive the osculating orbital equations:
\begin{subequations}
	\begin{equation}
		\PD{x_{\text{cons}}^\alpha}{I^A}  \frac{d I^A}{d t} = 0,
	\end{equation}
	\begin{equation}
		\PD{p^{\text{cons}}_\alpha}{I^A}  \frac{d I^A}{d t} = F_{\alpha},
	\end{equation}
\end{subequations}
For simplicity, we drop the ``cons'' subscript henceforth, but it is important to remember that all quantities in the above equations are evaluated on the tangent conservative orbit.
We also make use of the relationship:
\begin{equation}
	F_t = - \frac{1}{\mu}\frac{d H_{\text{EOB}}}{d t} = - \left( \frac{d r}{dt}F_r + \frac{d \phi}{d t} F_\phi \right). 
\end{equation}
With this in hand, using the second osculating condition gives us the following simultaneous equations:
\begin{subequations}
	\begin{equation}
	\PD{p_t}{p} \frac{d p}{d t} + \PD{p_t}{e} \frac{d e}{d t} = - \frac{d H_{\text{eff}}}{d H_{\text{EOB}}} \frac{d H_{\text{EOB}}}{d t} =  \frac{H_{\text{EOB}}}{M} F_t
	\end{equation}
	\begin{equation}
		\PD{p_\phi}{p} \frac{d p}{d t} + \PD{p_\phi}{e} \frac{d e}{d t} =  F_\phi,
	\end{equation}
\end{subequations}
which have the solution 
\begin{subequations} 
	\begin{align}
	\begin{split}
		\frac{d p}{d t} & = \frac{1}{D} \left( \PD{p_t}{e} F_\phi  -  \PD{p_\phi}{e} \frac{H_{\text{EOB}}}{M} F_t  \right),
	\end{split} \\
	\begin{split}
			\frac{d e}{d t} & = \frac{1}{D} \left( \PD{p_t}{p} F_\phi  -  \PD{p_\phi}{p} \frac{H_{\text{EOB}}}{M} F_t  \right), 
	\end{split} \\
	\begin{split}
			D & = \PD{p_t}{p} \PD{p_\phi}{e} -  \PD{p_t}{e} \PD{p_\phi}{p}.
	\end{split}
	\end{align}
\end{subequations}
 Here, $D$ corresponds to the Jacobian determinant of the transformation from $(H_\text{eff},p_\phi)$ to $(p,e)$, which is singular as $e \rightarrow 0$.

Using the radial component of the first osculating condition gives us
\begin{equation}
	\PD{r}{p} \frac{d p}{d t} + \PD{r}{e} \frac{d e}{d t} +  \PD{r}{w} \frac{d w}{d t}  = 0.
\end{equation}
Rearranging the above gives us
\begin{equation}
\label{eq:dwdt}
	\frac{d w}{d t}  = - \frac{1}{ \partial r / \partial w} \left( \PD{r}{p} \frac{d p}{d t} + \PD{r}{e} \frac{d e}{d t}   \right).
\end{equation}
Note that $\partial r / \partial w \rightarrow 0$ in the circular-orbit limit $e \rightarrow 0$, making the above expression singular. 
\footnote{This singularity and the singularity in the determinant of the Jacobian $D$ as $e \rightarrow 0$ can be circumvented through a different choice of orbital elements, namely $\alpha = e \sin w$ and $\beta = e \cos w$, which we will implement in a follow-up paper \cite{Lynch2026SmallEccentricities}.}

For planar orbits that we wish to orbit-average, it is far more convenient to parametrize the system with the radial phase parameter $\chi$ rather than with the time coordinate $t$ \cite{vandeMeent:2018cgn}. Using the chain rule, we find that
\begin{equation}
	\frac{d t}{d \chi} = \frac{d r / d \chi}{d r / dt } =  \frac{d r / d \chi}{ \xi(r) \partial H_\text{EOB} / \partial p_{r_*} }.
\end{equation}
Note that this expression yields $0/0$ at turning points where $d r / d \chi = d r / dt =0$. In the Schwarzschild limit, one can simplify this expression to remove these unphysical singularities, but this is not always possible in the EOB case. 

Moreover, rather than evolving $w$ on its own, we use $\zeta = \chi - w$ to absorb its evolution into that of the phase $\zeta$, thus reducing the orbital element parameter space by one.
Overall, we end up with equations of motion that take the following form:
\begin{subequations}
\begin{align}\label{eq:OOE_EoM}
	\begin{split}
		\frac{d p}{d \chi} &=  \frac{d p }{d t}  \frac{d t}{d \chi}  = \mr F_p (\nu; p,e,\zeta), 
	\end{split} \\
	\begin{split}
		\frac{d e}{d \chi} &=  \frac{d e }{d t}  \frac{d t}{d \chi}   = \mr F_e (\nu; p,e,\zeta) , 
	\end{split}\\
	\begin{split}
		\frac{d \zeta}{d \chi} &=  \frac{d \chi}{d \chi} -   \frac{d w }{d t}  \frac{d t}{d \chi}   = 1 + \mr  f_{w} (\nu; p,e,\zeta) , 
	\end{split} \\
	\begin{split}
		\frac{d t}{d \chi} & =  f_t  (\nu; p,e,\zeta)
	\end{split}\\
	\begin{split}
		\frac{d \phi}{d \chi} &=  \frac{d \phi }{d t} \frac{d t}{d \chi}  =   f_\phi   (\nu; p,e,\zeta)
	\end{split},
\end{align}
\end{subequations}
where the bookkeeping parameter $\mr$  marks which terms depend on radiation reaction.
We summarize osculating-orbital-element equations schematically as:
\begin{subequations} \label{eq:OOE_EoM_Eccentric_Param}
	\begin{align}
		\begin{split}
			& \dot{P_j} = \epsilon F_j (\nu,\vec{P}, \zeta) ,
		\end{split}\\
		\begin{split}
			& \dot{\zeta} = 1 + \epsilon f_{w}(\nu; \vec{P}, \zeta),
		\end{split}\\
		\begin{split}
			& \dot{S}_k = s_k(\nu;\vec{P}, \zeta),
		\end{split}
	\end{align}
\end{subequations}
where we denote the orbital elements $P_j = (p,e)$, we treat $\zeta$ as the radial phase, $S_k = (t,\phi)$ are the ``extrinsic" quantities, and an overdot denotes the rate of change with respect to $\chi$.
We stress that at this stage, the only approximation that has been made is for $p_{r*}(\nu;p,e,\zeta)$; otherwise these equations are exactly equivalent to the original EOB equations of motion.

\subsection{Regularization of forcing terms via Fourier decomposition} \label{section:FourierDecomposition}
The forcing terms $F_j$, $f_{w}$, and $s_k$ are all periodic functions of $\zeta$ with period $2 \pi$. However, they also have unphysical numerical singularities at the radial turning points where $\zeta = n \pi$ and $n \in \mathbb{Z}$. These singularities must be regularized before the equations of motion can be integrated numerically. We decompose functions into Fourier series using the convention:
\begin{equation} \label{eq:Fourier}
	A(\vec{P},\zeta) = \sum_{\kappa} A_{\kappa}(\vec{P}) e^{i  \kappa \zeta},
\end{equation}
where $ A $ represents any of the functions $F_j$, $f_{w}$, or $s_k$. 
Based on this, we can split the function into an averaged piece
\begin{equation}\label{eq:average}
	\avg{A} (\vec{P}) = A_{0}(\vec{P}) = \frac{1}{(2\pi)} \int_0^{2\pi} A(\vec{P},\zeta) d \zeta,
\end{equation}
and an oscillating piece
\begin{equation}\label{eq:oscillating}
	\osc{A}(\vec{P},\zeta) = A(\vec{P},\zeta) - \avg{A}(\vec{P})  = \sum_{\kappa \neq 0 } A_{\kappa}(\vec{P}) e^{i \kappa  \zeta}. 
\end{equation}

One way to compute the Fourier coefficients $A_\kappa(\vec{P})$ is to sample $A(\vec{P},\zeta)$ at equal steps in $\zeta$ over $[0,2 \pi ]$ and apply a fast Fourier transform (FFT). However, this requires evaluating the function at the turning points where it is singular. We therefore sample at the same steps but with a constant offset $\delta = 0.1$ to avoid the singular points. Using an FFT on this list gives us the coefficients $\hat{A}_\kappa(\vec{P})$, related to the original coefficients via 
\begin{equation}
	A_\kappa(\vec{P}) = \exp(i \kappa \delta) \hat{A}_\kappa(\vec{P}).
\end{equation}
We can reconstruct the original function $A(\vec{P},\zeta)$ via Eq.~\eqref{eq:Fourier} without ever evaluating it at the singular points.

Moreover, the Fourier coefficients decay exponentially with $|\kappa|$ so only a finite number of modes are needed to accurately reconstruct the function. For our work, we keep only $\kappa_\text{max} = 30$. Functions such as $f_t$ have Fourier modes that decay slowly with $\kappa$ but their reciprocal $f_t^{-1}$ decays much faster. We exploit this fact to accurately reconstruct $f_t$ using the method outlined in Appendix~\ref{section:ReciprocalFourierCoefficients}.

To obtain smooth functions over the parameter space, we numerically compute the Fourier coefficients for each of the forcing functions $F_j$, $f_{w}$, and $s_k$ on a grid in $\nu$, $p$, and $e$.
This yields regularized forcing functions evaluable anywhere in the parameter space, allowing us to integrate the osculating orbital element equations numerically. 

The resulting trajectories are much slower to compute than the EOB trajectories and do not match them perfectly owing to numerical errors from interpolation, Fourier-mode truncation, and the numerical ODE solver. Since we use the same interpolation grid, number of Fourier modes, and ODE solver throughout, comparing the OOE trajectories with the full EOB trajectories acts as a control for these errors and isolates the errors introduced by the near-identity averaging transformations of the next section.

\section{Near-Identity Averaging Transformations} \label{section:NITs}
Near-identity (averaging) transformations (NITs) are a well-known technique in applied mathematics and celestial mechanics \cite{Kevorkian1987}.
This technique involves making small transformations to the equations of motion, so that the short timescale physics is averaged out while the long-term evolution is retained. 
This is well suited to EMRIs, where we do not need the trajectory on the orbital timescale so long as we can accurately track its evolution on the radiation-reaction timescale \cite{vandeMeent:2018cgn,McCart:2021upc,Lynch:2021ogr,Pound2022,Lynch:2022zov,Lynch:2023gpu,Drummond:2023wqc,Lynch:2024ohd,Drummond:2026haw}.
We will show that NITs are also well suited to accelerating eccentric EOB inspiral dynamics.

%% NIT
We first define NIT variables $\nit{P}_j$, $\nit{\zeta}$, and $\nit{S}_k$, which are related to the OOE variables  $P_j$, $\zeta$ and $S_k$ via the following near-identity transformations:
\begin{subequations} \label{eq:transformation}
\begin{align}
	\begin{split}\label{eq:transformation1}
		\nit{P}_j &= P_j + \sp Y_j^{(1)}(\vec{P}, \zeta) + \sp^2 Y_j^{(2)}(\vec{P}, \zeta)  +  \HOT{3},
	\end{split}\\
	\begin{split}
		\nit{\zeta} &=  \zeta + \sp X^{(1)}(\vec{P}, \zeta) + \HOT{2},
	\end{split}\\
	\begin{split}
		\nit{S}_k &= S_k + Z_k^{(0)}(\vec{P}, \zeta) +\sp Z_k^{(1)}(\vec{P}, \zeta) + \HOT{2}.
	\end{split}
\end{align}
\end{subequations}
Note that the above functions also depend on the symmetric mass ratio $\nu$, which we suppress for notational convenience.
Here, the transformation functions $Y_j^{(n)}$, $X_i^{(n)}$, and $Z_k^{(n)}$ are required to be smooth, periodic functions of the radial phase $\zeta$. 
At leading order, Eqs.~\eqref{eq:transformation} are identity transformations for $P_k$ and $\zeta$ but not for $S_k$ owing to the zeroth order transformation term $Z_k^{(0)}$. 
%% Inverse NIT
The inverse transformations can be found for $P_j$ and $\zeta$ by requiring that their composition with the transformations in Eqs.~\eqref{eq:transformation} must recover the identity transformation. Expanding order by order in $\epsilon$, this gives us
\begin{subequations}\label{eq:inverse_transformation}
	\begin{align}
	\begin{split}
			P_j & = \nit{P_j} - \epsilon Y_j^{(1)}(\vec{\nit{P}},\nit{ \zeta}) - \epsilon^2 \nit{Y}_j^{(2)}(\vec{\nit{P}},\nit{ \zeta}) + \mathcal{O}(\epsilon^3),
	\end{split}\\
	\begin{split}
			\zeta  & = \nit{ \zeta} - \epsilon X^{(1)}(\vec{\nit{P}},\nit{ \zeta}) + \mathcal{O}(\epsilon^2),
	\end{split}\\
		\begin{split} \label{eq:inverse_transformation_S}
		S_k  & = \nit{S}_k -Z_k^{(0)}(\vec{\nit{P}},\nit{ \zeta}) - \epsilon \nit{Z}_k^{(1)}(\vec{\nit{P}},\nit{ \zeta}) + \mathcal{O}(\epsilon^2),
	\end{split}
\end{align}
\end{subequations}
where 
\begin{subequations}\label{eq:inverse_tramsformation_definitions}
	\begin{equation}
		\begin{split}
			\nit{Y}_j^{(2)} &= \left(  Y_j^{(2)} - \PD{Y_j^{(1)}}{\nit{P_k}} Y_k^{(1)} - \PD{Y_j^{(1)}}{\nit{ \zeta}} X^{(1)}  \right),
		\end{split}
	\end{equation}
	\begin{equation}
		\begin{split}
			\nit{Z}_k^{(1)} &= \left(  Z_k^{(1)} - \PD{Z_k^{(0)}}{\nit{P_k}} Y_k^{(1)} - \PD{Z_k^{(0)}}{\nit{ \zeta}} X^{(1)}  \right).
		\end{split}
	\end{equation}
\end{subequations}

% Summarize NIT results
To derive the averaged equations of motion, one takes the $\chi$ derivative of Eqs.~\eqref{eq:transformation}, substitutes the OOE equations of motion \eqref{eq:OOE_EoM_Eccentric_Param}, converts from OOE variables to NIT variables using Eqs.~\eqref{eq:inverse_transformation}, and expands order by order in $\mr$. At each order, one then fixes the form of the transformation functions $Y_j^{(n)}, X^{(n)}$ and $Z_k^{(n)}$ such that the resulting equations of motion for $\nit{P}_j, \nit{\zeta}$ and $\nit{S}_k$ take the following form
% NIT EoM
\begin{subequations}\label{eq:transformed_EoM}
	\begin{align}
	\begin{split}
		\dot{\nit{P}}_j & = \epsilon \nit{F}_j^{(0)}(\vec{\nit{P}}) + \epsilon^2 \nit{F}_j^{(1)}(\vec{\nit{P}}) + \epsilon^3 \nit{F}_j^{(2)}(\vec{\nit{P}}) +  \HOT{4},
	\end{split} \\
	\begin{split}
		\dot{\nit{\zeta}} & = 1 +\epsilon \nit{f}_{w}^{(1)}(\vec{\nit{P}}) +\epsilon^2 \nit{f}_{w}^{(2)}(\vec{\nit{P}})+  \HOT{3},
	\end{split}\\
	\begin{split}
		\dot{\nit{S}}_k & = \nit{s}_k^{(0)}(\vec{\nit{P}}) + \sp \nit{s}_k^{(1)}(\vec{\nit{P}}) + \sp^2  \nit{s}_k^{(2)}(\vec{\nit{P}})  +  \HOT{3}.
	\end{split}
\end{align}
\end{subequations}
where the superscripts in parentheses indicate the postadiabatic (PA) order of the terms.

Here we use "postadiabatic order" to refer to corrections to the adiabatic evolution of the orbital elements of an eccentric inspiral that arise from the NIT expansion. Leading order in the expansion corresponds to the adiabatic order (0PA), next-to-leading order corresponds to the first postadiabatic order (1PA), and next-to-next-to-leading order corresponds to the second postadiabatic order (2PA).

Our definition of postadiabatic orders most closely follows the literature of EMRIs and GSF theory \cite{Hinderer:2008dm} with one crucial distinction: we do not perform an expansion in the mass-ratio. As such, our usage differs slightly from that of GSF, where the postadiabatic order is also tied to the corresponding order of the self-force in the mass-ratio expansion. It also differs from the quasicircular postadiabatic expansion of Refs.~\cite{Nagar:2018gnk,Rettegno:2019tzh,Riemenschneider:2021ppj,Mihaylov:2021bpf}, which expands about the adiabatic evolution of a quasicircular inspiral.

Crucially, the above equations of motion are now independent of the orbital phase $\zeta$. In this procedure, one fixes only the oscillating parts of the transformation functions and is free to choose their orbit-averaged parts. Many such choices are explored in Ref.~\cite{vandeMeent:2018cgn}, but for simplicity we set all the orbit-averaged parts of the transformation functions to be zero. For brevity, we provide an ancillary Mathematica file that derives the relationships between ($\nit{F}_j^{(n)},\nit{f}_w^{(n)} $, and $\nit{s}_k^{(n)}$) to the terms in the OOE equations of motion and summarize the results below.

\subsection{Adiabatic terms}
The adiabatic (0PA) order averaged terms are simply given by
\begin{subequations}
	\begin{equation}
		\nit{F}_j^{(0)} = \left<F_{j}\right>,
	\end{equation}
	\begin{equation} \label{eq:0PA_f_t_f_phi}
		\nit{s}_k^{(0)} =  \left<s_{k}\right>.
	\end{equation}
\end{subequations}
At this order, the model reduces to the standard adiabatic model where one simply orbit-averages the forcing terms in the OOE equations of motion.
In deriving these relations we have constrained the oscillating pieces of the NIT transformation functions to be
\begin{subequations}
\begin{equation}\label{eq:qc_NIT_Y1}
	\osc{Y}_j^{(1)} = \sum_{\kappa \neq 0} \frac{i}{\kappa} F_{j,\kappa} e^{i \kappa \zeta},
\end{equation}
\begin{equation}\label{eq:qc_NIT_Z0}
	\osc{Z}_k^{(0)}  = \sum_{\kappa \neq 0} \frac{i}{\kappa} s_{k,\kappa}e^{i \kappa \zeta}.
\end{equation}
\end{subequations}
This leading order transformation between OOE and NIT variables is crucial for setting consistent initial conditions, ensuring we are comparing the same physical systems. Thus, we refer to the model that uses only the above 0PA equations of motion and leading order transformations for the initial conditions as the ``0PA model."

\subsection{First postadiabatic terms}
At first postadiabatic order (1PA), one obtains
\begin{subequations} \label{eq:1PA_Averaged_Terms_Ecc}
		\begin{equation}
		\nit{F}_j^{(1)} = \left<\frac{\partial \osc{Y}_j^{(1)}}{\partial \nit{P}_k} \osc{F}_k \right> + \left<\frac{\partial \osc{Y}_j^{(1)}}{\partial \nit{\zeta}} \osc{f}_{w} \right>,
	\end{equation}
		\begin{equation}
		\nit{f}_{w}^{(1)}= \left<f_{w}\right>,
	\end{equation}
	\begin{equation}
		\nit{s}_k^{(1)} =  \left<\frac{\partial \osc{Z}_k^{(0)}}{\partial \nit{P}_j} \osc{F}_j \right> + \left<\frac{\partial \osc{Z}_k^{(0)}}{\partial \nit{\zeta}} \osc{f}_{w} \right> ,
	\end{equation}
\end{subequations}
where we have constrained
\begin{subequations}
\begin{align}
	\begin{split}\label{eq:qc_NIT_Y2}
		\osc{Y}_j^{(2)}  = & \sum_{\kappa \neq 0} \frac{ i e^{ i \kappa \zeta}}{\kappa}  \Biggl[ -  \PD{\nit{F}_j^{(0)}}{\nit{P}_k} \osc{Y}_{k, \kappa}^{(1)} +   \\ 
		&  \sum_{\kappa' \neq 0} \left(  \frac{\partial \osc{Y}_{j, \kappa}^{(1)}}{\partial \nit{P}_k} \osc{F}_{k,\kappa - \kappa'}   +   \frac{\partial \osc{Y}_{j,\kappa'}^{(1)}}{\partial \nit{\zeta}} \osc{f}_{w,\kappa - \kappa'} \right) \Biggr]
	\end{split} \\
	\begin{split}\label{eq:NIT_X1}
			\osc{X}^{(1)}  = & \sum_{\kappa \neq 0} \frac{i}{\kappa} f_{w,\kappa}  e^{i \kappa \zeta},
	\end{split}\\
	\begin{split}\label{eq:qc_NIT_Z1}
	\osc{Z}_k^{(1)}  = & \sum_{\kappa \neq 0} \frac{ i e^{ i \kappa \zeta}}{\kappa}  \Biggl[ -  \PD{\nit{s}_k^{(0)}}{\nit{P}_j} \osc{Y}_{j, \kappa}^{(1)} +   \\ 
	&  \sum_{\kappa' \neq 0} \left(  \frac{\partial \osc{Z}_{k, \kappa}^{(0)}}{\partial \nit{P}_j} \osc{F}_{j,\kappa - \kappa'}   +   \frac{\partial \osc{Z}_{k,\kappa'}^{(0)}}{\partial \nit{\zeta}} \osc{f}_{w,\kappa - \kappa'} \right) \Biggr]
	\end{split}
\end{align}
\end{subequations}
We refer to the model using these 1PA equations of motion and together with the subleading transformations as the ``1PA model".

There is some ambiguity whether these subleading transformations $\osc{Y}_j^{(2)}$, $\osc{X}^{(1)}$, and $\osc{Z}_k^{(1)}$ belong at 1PA order, or whether only the leading order transformations should be included. We find that without them, the 1PA model is no more accurate than the 0PA model. We therefore choose to include them in our 1PA model.
\subsection{Second postadiabatic terms}
To reach acceptable accuracy for comparable-mass systems, we extend the calculation of the NIT equations of motion to second postadiabatic order (2PA). The resulting averaged terms are
\begin{subequations}
\begin{align} \label{eq:2PA_Averaged_Terms_Ecc}
	\begin{split}
		\nit{F}_j^{(2)} = & \left<\frac{\partial \osc{Y}_j^{(2)}}{\partial \nit{P}_k} \osc{F}_{k} \right> + \left<\frac{\partial \osc{Y}_j^{(2)}}{\partial \nit{\zeta}} \osc{f}_{w} \right> 
		\\ & -  \frac{1}{2} \frac{\partial^2 \nit{F}_j^{(0)}}{\partial \nit{P}_k \partial \nit{P}_i}  \left< \osc{Y}_k^{(1)} \osc{ Y}_i^{(1)} \right> ,
	\end{split}\\
	\begin{split}
		\nit{f}_{w}^{(2)} = & \left<\PD{\osc{X}^{(1)} }{\nit{P}_j} F_j \right> +  \left<\PD{\osc{X}^{(1)} }{\nit{\zeta}} f_{w}  \right>,
	\end{split}\\
	\begin{split}
		\nit{s}_k^{(2)} = & \left<\frac{\partial \osc{Z}_k^{(1)}}{\partial \nit{P}_j} \osc{F}_{j} \right> + \left<\frac{\partial \osc{Z}_k^{(1)}}{\partial \nit{\zeta}} \osc{f}_{w} \right> 
		\\ & - \frac{1}{2} \frac{\partial^2 \nit{s}_k^{(0)}}{\partial \nit{P}_i \partial \nit{P}_j}  \left< \osc{Y}_i^{(1)} \osc{ Y}_j^{(1)} \right> .
	\end{split}
\end{align}
\end{subequations}
We refer to the model that adds these 2PA contributions to the 1PA as the ``2PA model". Any accuracy gain from 1PA to 2PA can be attributed to these terms.

At this order, the equations of motion become ill-behaved in the circular-orbit limit $e \rightarrow 0 $ where $f_{w}$ diverges as $1/e$. Moreover, the osculating-orbit eccentricity ceases to behave as a smooth orbit-averaged evolution plus a rapidly oscillating contribution. This phenomenon will be explored in a follow-up paper \cite{Lynch2026SmallEccentricities}.

\subsection{Summary of results}
Combining the above results, we obtain the averaged equations of motion for the NIT variables through second postadiabatic order:
\begin{subequations}
\begin{align}
	\begin{split}
		\frac{d \nit{p}}{d \chi} &=  \epsilon \nit{F}_p^{(0)}(\nit{p},\nit{e}) +  \epsilon^2 \nit{F}_p^{(1)}(\nit{p},\nit{e}) +  \epsilon^3 \nit{F}_p^{(2)}(\nit{p},\nit{e}) + \HOT{4} , 
	\end{split} \\
	\begin{split}
		\frac{d \nit{e}}{d \chi} &=  \epsilon \nit{F}_e^{(0)}(\nit{p},\nit{e}) +  \epsilon^2 \nit{F}_e^{(1)}(\nit{p},\nit{e}) +  \epsilon^3 \nit{F}_e^{(2)}(\nit{p},\nit{e}) + \HOT{4}, 
	\end{split}\\
	\begin{split}
		\frac{d \nit{\zeta}}{d \chi} &=1 +  \epsilon  \nit{f}_{w}^{(1)}(\nit{p},\nit{e}) + \epsilon^2 \nit{f}_{w}^{(2)} (\nit{p},\nit{e}) + \HOT{3}, 
	\end{split} \\
	\begin{split}
		\frac{d \nit{t}}{d \chi} & =  f_t^{(0)} (\nit{p},\nit{e}) +  \epsilon  \nit{f}_t^{(1)} (\nit{p},\nit{e})  +  \epsilon^2 \nit{f}_t^{(2)} (\nit{p},\nit{e}) + \HOT{3}
	\end{split}\\
	\begin{split}
		\frac{d \nit{\phi}}{d \chi} & =  f_\phi^{(0)} (\nit{p},\nit{e}) +  \epsilon \nit{f}_\phi^{(1)} (\nit{p},\nit{e}) +  \epsilon^2 \nit{f}_\phi^{(2)} (\nit{p},\nit{e}) + \HOT{3}
	\end{split}
\end{align}
\end{subequations}
One can then relate the NIT variables to the OOE variables to ensure consistent initial conditions via
\begin{subequations}
\begin{align} \label{eq:OOE_NIT_Initial_Conditions}
	\begin{split}
		\nit{p} &= p + \epsilon Y_p^{(1)}(p,e,\zeta ) +  \epsilon^2 Y_p^{(2)}(p,e,\zeta ) + \HOT{3} , 
	\end{split} \\
	\begin{split}
		\nit{e} &=  e + \epsilon Y_e^{(1)}(p,e,\zeta ) +  \epsilon^2 Y_e^{(2)}(p,e,\zeta ) + \HOT{3}
	\end{split}\\
	\begin{split}
		\nit{\zeta} &= \zeta +  \epsilon  X^{(1)}(p,e,\zeta ) + \HOT{2},
	\end{split} \\
	\begin{split}
		\nit{t} & = t +  Z_t^{(0)} (p,e,\zeta ) +  \epsilon  Z_t^{(1)} (p,e,\zeta )  +  \HOT{2}
	\end{split}\\
	\begin{split}
		\nit{\phi} & = \phi +  Z_\phi^{(0)} (p,e,\zeta ) +  \epsilon  Z_\phi^{(1)} (p,e,\zeta )  +  \HOT{2}
	\end{split}
\end{align}
\end{subequations}
The OOE variables can then be recovered from the solutions to the NIT equations of motion using the inverse transformations of Eqs.~\eqref{eq:inverse_transformation}.
A \textsc{Mathematica} notebook that carries out the derivation of this section, including the inverse transformation and the resulting evolution of the orbital elements, phases, and extrinsic quantities, is provided as Supplemental Material \cite{SupplementalMaterial}.

\section{Waveform Generation} \label{section:Waveforms}
We now discuss how to generate waveforms from both the EOB and averaged dynamics.
The waveform strain $h$ can be split into polarizations $h_+$ and $h_\times$, which can be further decomposed into $-2$ spin-weighted spherical-harmonic modes as follows:
\begin{equation} \label{eq:StrainDecomposition}
	h = h_+ - i h_\times = \sum_{\ell \geq 2} \sum_{m=-\ell}^{\ell} h_{\ell m} \,_{-2}Y_{\ell m}(\theta, \phi).
\end{equation}
Within the PN formalism, the waveform modes $h_{\ell m}$ have been calculated up to 3PN order \cite{Mishra:2015bqa,Boetzel:2019nfw, Ebersold:2019kdc,Henry:2023tka}, and transformed into the EOB formalism in Ref.~\cite{Gamboa:2024imd}.
As a proof of principle, here we employ the 3PN instantaneous contributions to the modes from Ref.~\cite{Gamboa:2024imd}, which take the general form:
\begin{equation}
	h_{\ell m} =  \mathcal{A}_{\ell m} \left(r,p_r,p_\phi \right) \exp \left( - i m \phi \right).
\end{equation}
We model only the dominant $(\ell,m) = (2,2)$ mode, so we compute only $h_{22}$. We provide explicit expressions for the $\mathcal{A}_{22}$ mode amplitudes up to 3PN order for nonspinning systems in Appendix~\ref{section:WaveformModes}.

\subsection{Waveforms from EOB dynamics}

To generate waveforms from the EOB dynamics, one first solves the EOB equations of motion for the variables $(r(t),p_{r_*}(t), \phi(t), p_\phi(t))$ using standard numerical ODE solvers such as adaptive step-size Runge-Kutta methods.

Once solved, the EOB dynamics yield the solutions $(r(t),p_{r_*}(t), \phi(t), p_\phi(t))$ on nonuniform time steps. At each time step, one substitutes these solutions into the waveform-mode expressions, here $h_{2 2}$.
Finally, the modes are interpolated onto a dense uniformly sampled time grid that can then be Fourier transformed for mismatch and likelihood calculations.

\subsection{Waveforms from NIT dynamics} \label{section:WaveformsNIT}
When using the NIT equations of motion, one obtains solutions for the NIT variables $(\nit{p}(\chi), \nit{e}(\chi), \nit{\zeta}(\chi), \nit{t}(\chi), \nit{\phi}(\chi))$ on sparse, nonuniform steps in $\chi$.     Substituted directly into the waveform mode expressions, these are too sparse to resolve the orbital-timescale oscillations of the waveform. We therefore first sample the NIT solutions onto a uniform grid in $\chi$, with spacing $\Delta \chi$ small enough to resolve the orbital-timescale oscillations as set by the Nyquist theorem:
\begin{equation} \label{eq:Nyquest}
	\Delta \chi \leq \frac{2 \pi}{2 \kappa_\text{step} + 1}.
\end{equation}
Here $\kappa_\text{step}$ is the highest Fourier mode present in the inverse transformation for the extrinsic NIT variable $t$ given by Eq.~\eqref{eq:inverse_transformation_S}. We determine this by examining the ratio of the Fourier coefficients of the leading transformation until we find 
\begin{equation}
	\left| \frac{Z_{t,\kappa_\text{step}}^{(0)}(p_0,e_0) }{ Z_{t,1}^{(0)} (p_0,e_0)} \right|  \leq 10^{-16},
\end{equation}
as any smaller mode is of the same order of magnitude as machine precision. To ensure we never undersample our waveforms at lower eccentricities, we take the maximum of the above criterion and $\kappa_\text{step} = 20$.
Thus for lower eccentricities, we can sample at lower rates than for higher eccentricity systems. This condition can be relaxed to larger values of $\Delta \chi$, further reducing waveform-generation cost at some loss of accuracy. However, since our priority is to test the accuracy of the averaged dynamics, we choose a conservative sampling criterion.

In principle, one should also transform from the intrinsic NIT variables $(\nit{p}, \nit{e}, \nit{\zeta})$ to the osculating orbital element variables $(p, e, \zeta)$ on this uniform $\chi$ grid using Eqs.~\eqref{eq:inverse_transformation}. Even for the worst-case scenario of equal-mass, large eccentricity inspirals, this approximation only ever results in a mismatch of $\sim 10^{-6}$ and so can safely be implemented, just like in the extreme-mass-ratio case \cite{vandeMeent:2018cgn,Lynch:2021ogr,McCart:2021upc,Lynch:2022zov,Lynch:2023gpu,Lynch:2024ohd}.  Thus one can reconstruct the EOB variables $(r(\chi), p_{r_*}(\chi), \phi(\chi), p_\phi(\chi))$ using the relations of Sec.~\ref{section:QuasiKeplerianParametrization} with the approximation $(p, e, \zeta) \sim (\nit{p}, \nit{e}, \nit{\zeta})$. Using the mode expressions in Appendix~\ref{section:WaveformModes} one obtains the waveform on a nonuniform time grid. Finally, this is interpolated and upsampled onto a dense uniform time grid that can then be Fourier transformed for mismatch and likelihood calculations.

\section{Implementation}\label{section:Implementation}
We now outline the implementation of our Mathematica code for solving the dynamics and generating waveforms. This covers the interpolation grid, the initial conditions, the numerical integration, and the handover conditions for switching from NIT to EOB dynamics. We also describe how we compute mismatches between NIT/OOE and EOB waveforms.

\subsection{Interpolation grid} \label{section:InterpolationGrid}
To obtain smooth functions over the parameter space, we numerically calculate the Fourier coefficients via a fast Fourier transform (FFT) for each of the forcing functions $F_j$, $f_{w}$, and $s_k$  on a grid in $\nu$, $p$, and $e$. Motivated by the fact that the rate of change of $p$ with respect to time scales as $p^{-3}$ in the weak field limit \cite{Peters:1964zz}, we define the quantity
	$u = \log \left( p^{-3} \right)$.
An equally spaced grid in $u$ covers both the strong- and weak-field regimes with sufficient resolution.

We then construct a grid that has uniform spacing in $\nu$, $e$, and $u$ given by
\begin{itemize}
	\item $\nu \in [0.01,0.25 ]$ with $\Delta \nu = 0.01$,
	\item $e \in [ 0.01, 0.9]$ with $\Delta e = 0.01$,
	\item $u \in [ -17, \log \left((6 + 2 e_\text{max})^{-3}\right)]$ (which corresponds to $p \in [6 + 2 e_\text{max}, 289.069 ]$) with $\Delta u = 0.05$.
\end{itemize}
This results in a grid of $25 \times 90 \times 217 = 488,250$ points. We then use cubic-spline interpolation to evaluate the Fourier coefficients at any point in the parameter space.

For the postadiabatic transformation functions and averaged terms, we combine the Fourier coefficients of the OOE forcing functions with the relationships derived in Sec.~\ref{section:NITs} on the same grid, and then use cubic-spline interpolation to evaluate them rapidly at any point in the $(\nu,p,e)$ parameter space.

\subsection{Initial conditions} \label{section:InitialConditions}
To facilitate comparisons between the EOB, OOE, and NIT dynamics, we must specify consistent initial conditions for each system. 
The initial conditions for the OOE system are $(p_0, e_0, \zeta_0, \phi_0)$ at an initial time $t_0$. This corresponds to the purely conservative orbit to which the inspiral is instantaneously tangential. Often rather than $p_0$, we specify the initial orbit-averaged waveform frequency $f_0$, and root-find numerically for the value of $p_0$ that satisfies
\begin{equation}
	f_0 =  \frac{2}{2 \pi} \avg{\Omega_\phi} = \frac{1}{\pi} \left( \frac{\nit{f}_\phi^{(0)}(\nu,p_0,e_0)}{\nit{f}_t^{(0)}(\nu,p_0,e_0)} \right).
\end{equation}
Here $\nit{f}_\phi^{(0)}$ and $\nit{f}_t^{(0)}$ are the leading order terms in the NIT equations of motion for the extrinsic variables $\phi$ and $t$, respectively as given by Eq.~\eqref{eq:0PA_f_t_f_phi}. 

To find the corresponding initial conditions for the EOB system $(r_0, p_{r_* 0}, \phi_0, p_{\phi0})$, one can use the relations Eq.~\eqref{eq:r_of_p_e_xi}, Eq.~\eqref{eq:pphi_of_p_e}, and the relation for the initial radial momentum, given in Appendix \ref{section:prCoefficients}.\footnote{Note that this differs from the approaches used in Ref.~\cite{Ramos-Buades:2021adz,Gamboa:2024hli,Gamboa:2026jht} where one also accounts for the effect of radiation reaction on the initial conditions for the EOB system for consistency with quasi-circular \texttt{SEOBNR} models}

To set the initial conditions for the NIT system, we use Eqs.~\eqref{eq:OOE_NIT_Initial_Conditions} to transform the OOE initial conditions into the corresponding NIT ones. We use only the leading order terms for the 0PA model and the subleading terms for higher PA models.
This ensures that the initial conditions of all systems are consistent and correspond to the same physical inspiral, enabling meaningful comparisons between the resulting waveforms.

\subsection{Numerical integrator} \label{section:NumericalIntegrator}
To solve the equations of motion, we use the \texttt{NDSolve} function in Mathematica with the method set to \texttt{ExplicitRungeKutta} with \texttt{AccuracyGoal} set to $9$ and \texttt{PrecisionGoal} set to $\infty$ to ensure that numerical errors from the ODE solver do not dominate the phase error for the long inspiral trajectories computed here. For the EOB equations of motion, we set the \texttt{DifferenceOrder} to $4$ (i.e., RK4) to ensure numerical stability for long, eccentric inspirals. For the OOE and NIT equations of motion, the system of equations is much less stiff allowing \texttt{DifferenceOrder} $8$ (i.e., RK8), and hence, larger time steps and faster run times. When transitioning back to the EOB dynamics for plunge, we again use RK4 for stability.

\subsection{Handover conditions} \label{section:Handover}
The OOE and NIT equations of motion are valid only during the inspiral and break down as the system approaches plunge and merger. We therefore specify handover conditions for switching from NIT to EOB dynamics to model the late inspiral and plunge accurately.

First we use the adiabaticity parameter given by $ \mathcal{Q} = \avg{\Omega_\phi}^2 / \avg{\dot{\Omega}_\phi} \gg 1$. We find empirically that the NIT equations of motion become significantly less accurate when $\mathcal{Q} \lesssim 100$, though the resulting waveform accuracy is not highly sensitive to this exact threshold. To evaluate this quickly at each time step, we use the following approximation
\begin{equation}
\begin{aligned}
	\mathcal{Q} & \sim \frac{\left(\nit{f}^{(0)}_\phi / \nit{f}^{(0)}_t \right)^2} { \frac{d}{dt} \left(\nit{f}^{(0)}_\phi / \nit{f}^{(0)}_t \right)} 
	\\& \sim \left(\nit{f}^{(0)}_\phi \right)^2 \nit{f}^{(0)}_t \left[ \left( \PD{\nit{f}^{(0)}_\phi}{\nit{P}_j} \nit{f}^{(0)}_t  - \PD{\nit{f}^{(0)}_t}{\nit{P}_j} \nit{f}^{(0)}_\phi  \right) \nit{F}_j^{(0)} \right]^{-1}.
\end{aligned}
\end{equation}

This condition is too permissive for low-eccentricity, large-mass-ratio systems, and so we add another handover condition when $\tilde{p} = 8.5$.

Finally, we also impose a handover when $\tilde{e} < 0.035$ since the NIT equations become ill-behaved as $e \rightarrow 0 $ where $f_{w}$ diverges as $1/e$. This condition will be replaced with a transformation to NIT variables that remain well behaved in the small-eccentricity limit, derived in a follow-up paper \cite{Lynch2026SmallEccentricities}.

In summary, when evolving the NIT equations of motion, we check at each time step if any of the following conditions are violated:
\begin{itemize}
	\item $\tilde{p} < 8.5$,
	\item $\mathcal{Q} < 100$,
	\item $\tilde{e} < 0.035$.
\end{itemize}
If any of these is violated, we switch from the NIT to the EOB equations of motion. We have chosen these conditions conservatively to prioritize the accuracy of the NIT waveforms compared to pure EOB waveforms. More optimal handover conditions undoubtedly exist that retain this accuracy while allowing a longer NIT evolution, but we leave such optimization to future work.

\subsection{Mismatch calculation} \label{section:MismatchCalculation}

To quantify the accuracy of NIT and OOE waveforms relative to EOB waveforms, we require a measure of similarity between them. The standard measure used in GW data analysis is the noise-weighted inner product between two waveforms $h_1$ and $h_2$,
\begin{equation}
	\left<h_1 | h_2 \right> = 4 \Re \int_{f_\text{low}}^{f_\text{high}} \frac{\tilde{h}_1(f) \tilde{h}_2^*(f)}{S_n(f)} df,
\end{equation}
where $\tilde{h}(f)$ is the Fourier transform of the waveform $h(t)$, $\tilde{h}^*(f)$ is the complex conjugate of $\tilde{h}(f)$, $S_n(f)$ is the one-sided power spectral density (PSD) of the detector noise, and $f_\text{low}$ and $f_\text{high}$ are the lower and upper frequency cutoffs \cite{Finn:1992wt}.

To represent current ground-based detectors, we use the \texttt{A+} LIGO PSD with a low-frequency cutoff of 10 Hz, corresponding to the design sensitivity for the fifth observing run (O5)~\cite{LIGO-T0900288-v3}. We also use the \texttt{EinsteinTelescope} and \texttt{CosmicExplorer2} PSDs from Ref.~\cite{LIGO-T0900288-v3} with a low-frequency cutoff of 5 Hz. For LISA, we use an analytic estimate of the design sensitivity curve with a galactic confusion noise for the nominal four-year mission lifetime and a low-frequency cutoff of $10^{-4}$ Hz as given by Eq.~(1) of Ref.~\cite{Robson:2018ifk}. We use this PSD only to gain intuition into the dependence of the mismatch with expected frequency sensitivity of LISA. A rigorous assessment of the accuracy required for LISA would necessitate computing the full instrument response and projection onto TDI channels which can be explored in future work.

The inner product is then used to define the faithfulness between two waveforms as
\begin{equation}
	\mathcal{F} = \max_{t_c, \phi_c} \frac{\left<h_1 | h_2 \right>}{\sqrt{\left<h_1 | h_1 \right> \left<h_2 | h_2 \right>}},
\end{equation}
where the maximization over time and phase shifts $t_c$ and $\phi_c$ accounts for the fact that the two waveforms may not be perfectly aligned in time or phase \cite{Owen:1995tm}. The mismatch is then defined as
\begin{equation}
	\mathcal{M} = 1 - \mathcal{F}.
\end{equation}
Since we compute only the $(\ell,m) = (2,2)$ mode of the waveform, we report $\mathcal{M}_{22}$ for the EOB, OOE, and NIT waveforms. While the subdominant modes do not become significantly more important with eccentricity, they do for large-mass-ratio systems (see Fig.~11 of Ref.~\cite{Islam:2026blk}, Fig.~14 of Ref.~\cite{Gamboa:2026jht}, or Fig.~16 of~\cite{Faggioli:2025hff}).
However, since we are primarily interested in the accuracy of the NIT dynamics rather than the physical accuracy of the waveforms, 
this is an acceptable measure here.

For simulations of stellar-mass BBHs as observed by ground-based detectors, we use a sample rate of 4096 Hz, which gives us a high frequency resolution up to 2048 Hz via the Nyquist theorem. Since we exclude the merger and ringdown where the GW frequency sharply rises, this is more than sufficient to avoid aliasing. For simulations of massive black-hole binaries (MBHBs) as observed by LISA, we use a sample rate of 0.2 Hz ($\Delta t = 5s$), allowing us to resolve frequencies up to 0.1 Hz. 

For the mismatch calculations, we developed a Mathematica package \texttt{GWMMatch} \cite{lynch_2026_20426391} to compute time- and phase- optimized mismatches using the same methods as \texttt{PyCBC} \cite{pycbc_software}. Since we compare inspiral-only waveforms (without the merger and ringdown), we apply a Planck taper to both ends of the waveform to mitigate spectral leakage \cite{McKechan:2010kp}. We then zero-pad the waveforms to a common length, FFT them, and compute the noise-weighted inner product with the PSDs described above. Finally, we maximize over time and phase to find the faithfulness, and thus the mismatch. We use this mismatch to quantify the accuracy of NIT and OOE waveforms relative to EOB waveforms in the following sections.

\section{Results} \label{section:Results}

We now present the results from our implementation of NITs for eccentric EOB inspirals. We begin with a case study comparing EOB, OOE and NIT waveforms for a specific set of parameters. We then compare the accuracy of different postadiabatic orders across parameter space for total mass $M = 60M_\odot$ and then more closely examine the accuracy of the 2PA dynamics for several different total masses. Finally, we examine the speedup from the NIT dynamics and how it varies with both mass ratio and total mass.

\subsection{Case study} \label{section:CaseStudy}

% We select a specfic set of parameters.
We first demonstrate the accuracy of the NIT dynamics and waveform construction for a specific set of parameters. We choose a system with total mass $M = 20 M_\odot$, mass ratio $q = 1$, and initial eccentricity $e_0 = 0.75$ at an orbit-averaged reference frequency of $5$ Hz, corresponding to an initial semilatus rectum of $p_0 = 33.92$. This system evolves to an eccentricity $e \sim 0.625 $ at 10 Hz. We evolve to $r = 4 M$, where we terminate the dynamics and the waveform. Note that the innermost stable circular orbit for an equal-mass binary with our choice of effective Hamiltonian is at $r \simeq 4.45M$, so the final segment always corresponds to a plunging trajectory.

By rescaling the waveform with the total mass, we also examine the case of an MBHB system with $M = 10^6 M_\odot$ at the same mass ratio and initial eccentricity with a reference frequency of $10^{-4}$ Hz. 

In each case we generate EOB, OOE, and NIT (0PA,1PA,2PA) waveforms and compute mismatches between them for O5, CE, ET and LISA using the method described in Sec.~\ref{section:MismatchCalculation}.

\begin{figure*}
\includegraphics[width=\textwidth]{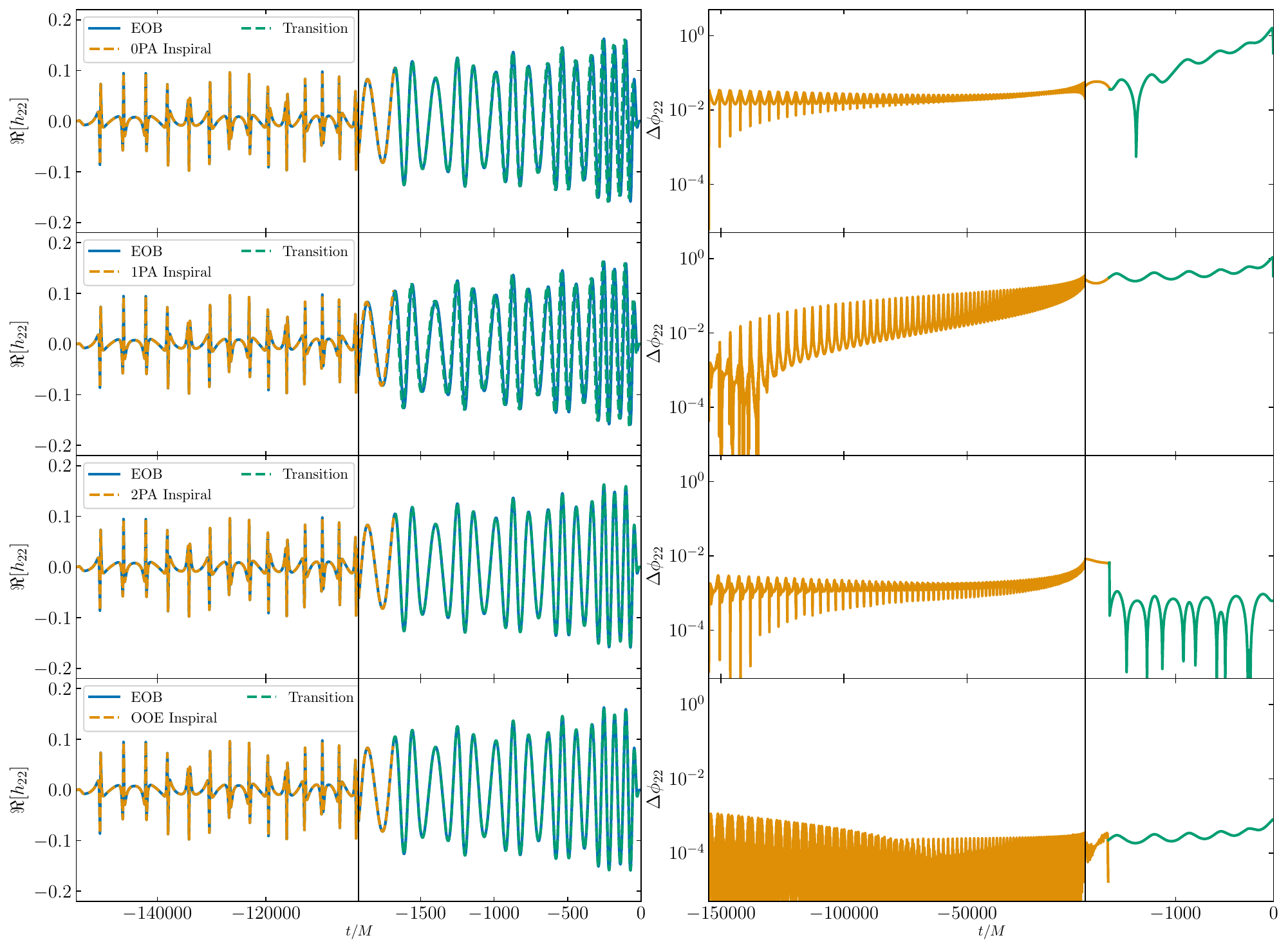}
\caption{The $(\ell,m) = (2,2)$ mode of the EOB, OOE and NIT waveforms (at various PA orders) for the case-study parameters $(q,p_0,e_0,\zeta_0) = (1,33.92,0.75,0)$ (orbit-averaged frequency of $5$ Hz) evolved until $r = 4 M$. The left panels show the early and late inspiral, highlighting where the NIT dynamics (dashed orange) transition to the EOB dynamics (dashed green). The right panels show the $(\ell,m) = (2,2)$ phase difference between the EOB and OOE/NIT waveforms at each PA order accumulated throughout the inspiral.
\label{fig:Waveform_Comparison}
}
\end{figure*}

In Fig.~\ref{fig:Waveform_Comparison}, we compare the EOB, OOE and NIT waveforms for the case study parameters. The left panels show the start and end of the inspiral, where one can see where we swap back from the NIT or OOE dynamics (dashed orange) to the EOB dynamics (dashed green). The right panels show the accumulated phase difference of the $(\ell,m) = (2,2)$ mode throughout the inspiral. The 0PA and 1PA NIT waveforms accumulate a significant phase difference of $\sim 1$ radian by $r = 4M$. In contrast, both the 2PA and OOE waveforms are visually indistinguishable from the EOB waveform and accumulate a phase difference of less than $10^{-2}$ radians by the end of the inspiral.

\begin{table*}%[H] add [H] placement to break table across pages
	\caption{\label{tab:CaseStudyMismatches} Timings and the $(2,2)$ mode mismatches $\mathcal{M}_{22}$ between EOB, NIT, and OOE waveforms for a BBH with $(M,f_0,e_{0}) = (20 M_\odot, 5 \text{Hz}, 0.75)$ [for the LISA case an MBHB with $(M,f_0,e_{0}) = (10^6 M_\odot, 10^{-4} \text{Hz}, 0.75)$], for mass ratios $q = (1,3,10)$, and for different detector PSDs (O5, CE, ET, and LISA).}
		\begin{tabular}{|c  c c c  c c c  c c|}
			 \hline
			 \textbf{Model} & \textbf{Dynamics [s]} & \textbf{Speedup} &\makecell{\textbf{Dynamics $\&$} \\ \textbf{Waveform [s]}} & \textbf{Speedup} &\textbf{O5} &\textbf{CE }&\textbf{ET }  &\textbf{LISA } \\
			 \hline 
			 $\mathbf{q = 1}$ & & & & & & & & \\
			 %\hline
			 EOB & $3.34$ & $\dots$ & $17.01$ & $\dots$ &$\dots$ & $\dots$ & $\dots$ & $\dots$  \\
			 %\hline
			NIT 0PA & $0.16$ & $20.69$ & $2.12$ & $8.04$ & $7.34 \times 10^{-3}$ &  $4.29 \times 10^{-3}$ &$8.07 \times 10^{-3}$ &  $8.09 \times 10^{-3}$  \\
			%\hline
			NIT 1PA & $0.25$ &$13.53$ & $2.33$ &$7.30$ & $6.55 \times 10^{-4}$ & $5.37 \times 10^{-4}$ & $6.02 \times 10^{-4}$ &  $1.44 \times 10^{-4}$  \\
			%\hline
			NIT 2PA & $0.24$ &$14.08$ & $2.63$ &$6.46$ & $1.64 \times 10^{-6}$ & $1.11 \times 10^{-6}$ & $1.47 \times 10^{-6}$ &  $7.45 \times 10^{-7}$  \\
			%\hline
			OOE & $15.68$ &$0.213$ & $17.27$ &$0.985$ & $1.50 \times 10^{-8}$ & $8.50 \times 10^{-9}$  & $1.49 \times 10^{-8}$ &  $8.91 \times 10^{-8}$  \\
			%\hline
			$\mathbf{q = 3}$ & & & & & & & & \\
			 %\hline
			 EOB & $4.31$ & $\dots$ & $22.74$ & $\dots$ &$\dots$ & $\dots$ & $\dots$ & $\dots$  \\
			 %\hline
			NIT 0PA & $0.16$ & $26.34$ & $2.86$ & $7.96$ & $8.96 \times 10^{-4}$ &  $3.83 \times 10^{-4}$ &$8.46 \times 10^{-4}$ &  $2.66 \times 10^{-3}$  \\
			%\hline
			NIT 1PA & $0.24$ &$17.66$ & $3.07$ &$7.40$ & $3.99 \times 10^{-4}$ & $3.29 \times 10^{-4}$ & $3.74 \times 10^{-3}$ &  $1.40 \times 10^{-4}$  \\
			%\hline
			NIT 2PA & $0.24$ &$18.07$ & $3.38$ &$6.73$ & $1.69 \times 10^{-6}$ & $9.82 \times 10^{-7}$ & $1.51 \times 10^{-6}$ &  $1.15 \times 10^{-6}$  \\
			%\hline
			OOE & $21.54$ &$0.20$ & $23.60$ &$0.96$ & $1.56 \times 10^{-8}$ & $8.93 \times 10^{-9}$  & $1.55 \times 10^{-8}$ &  $1.01 \times 10^{-7}$  \\
			%\hline 
			$\mathbf{q = 10}$ & & & & & & & & \\
			%\hline
			 EOB & $9.24$ & $\dots$ & $52.49$ & $\dots$ &$\dots$ & $\dots$ & $\dots$ & $\dots$  \\
			%\hline
			NIT 0PA & $0.19$ & $49.61$ & $6.34$ & $8.27$ & $3.31 \times 10^{-4}$ &  $9.08 \times 10^{-5}$ &$2.74 \times 10^{-4}$ &  $1.98 \times 10^{-3}$  \\
			%\hline
			NIT 1PA & $0.26$ &$35.18$ & $7.05$ &$7.45$ & $1.02 \times 10^{-4}$ & $8.25 \times 10^{-5}$ & $1.02 \times 10^{-4}$ &  $1.31 \times 10^{-4}$  \\
			%\hline
			NIT 2PA & $0.23$ &$40.81$ & $7.24$ &$7.25$ & $1.19 \times 10^{-6}$ & $4.57 \times 10^{-7}$ & $9.85 \times 10^{-7}$ &  $1.60 \times 10^{-6}$  \\
			%\hline
			OOE & $46.29$ &$0.20$ & $51.02$ &$1.03$ & $1.86 \times 10^{-8}$ & $1.11 \times 10^{-8}$  & $1.86 \times 10^{-8}$ &  $1.28 \times 10^{-7}$  \\
			\hline
		\end{tabular}
 \end{table*}

 We further test the NIT waveforms by computing their mismatches against EOB waveforms for each postadiabatic order for the case study parameters for different mass ratios $q = (1,3,10)$. We also report the walltimes for the dynamics alone and for the full waveform generation time. The results are shown in Table~\ref{tab:CaseStudyMismatches}.

 First, although the mismatches vary somewhat between detector PSDs, they agree to within an order of magnitude. Conclusions drawn from the O5 PSD should therefore be broadly applicable to other detectors as well.

 Second, the NIT dynamics (at all postadiabatic orders) speed up the dynamics by $\sim 10 - 50 \times$, scaling with mass ratio. The total waveform generation speedup is $\sim 6 - 8 \times$. It is smaller than the speedup of the dynamics alone because the dynamics are no longer the bottleneck - waveform generation is. This is underscored by the OOE waveforms, whose total walltime is comparable to that of the EOB waveforms despite the OOE dynamics being much slower. Our new method for sampling the waveform is much more efficient than the conventional method. We will explore how interpolating the waveform modes in terms of the NIT variables can further accelerate waveform generation and reduce this bottleneck in a follow-up paper \cite{Lynch2026SmallEccentricities}.

 Third, the 0PA and 1PA NIT waveforms have mismatches of $\mathcal{M}_{22} \sim 10^{-3}$ to $10^{-4}$, which decrease with mass ratio, as expected, since this increases the adiabaticity of the inspiral. The 2PA NIT waveforms reach $\mathcal{M}_{22} \sim 10^{-6}$ to $10^{-7}$ making 2PA the preferred order for comparable-mass systems. As a control, we include the OOE waveforms which quantify the accuracy lost by errors arising from interpolation, the ODE solver, and our method for sampling the waveform described in Sec.\ref{section:WaveformsNIT}. Their mismatches of $\mathcal{M}_{22} \sim 10^{-8}$ indicate that the NIT mismatches are dominated by the order at which we truncate the NIT expansion. One could push to higher PA orders to achieve further accuracy. A conservative estimate for the SNR at which the EOB and 2PA NIT waveforms would be distinguishable is given by $\sim 1 / \sqrt{2 \mathcal{M}}$ \cite{Lindblom:2008cm}. For $\mathcal{M}_{22} \sim 10^{-6}$, this would be an SNR of $\sim 700$. Thus going beyond 2PA order would be unnecessary for most comparable mass binaries, bar exceptionally loud MBHB signals that one might observe with LISA \cite{Li_2022,LISA:2024hlh,LISAConsortiumWaveformWorkingGroup:2023arg}. 

 In the following sections, we will see that these results broadly hold across the eccentric BBH parameter space for current ground-based detectors.

\subsection{Comparison of different postadiabatic orders} \label{section:PAComparison}

\begin{figure}[tbp]
	\includegraphics[width=1.0\columnwidth]{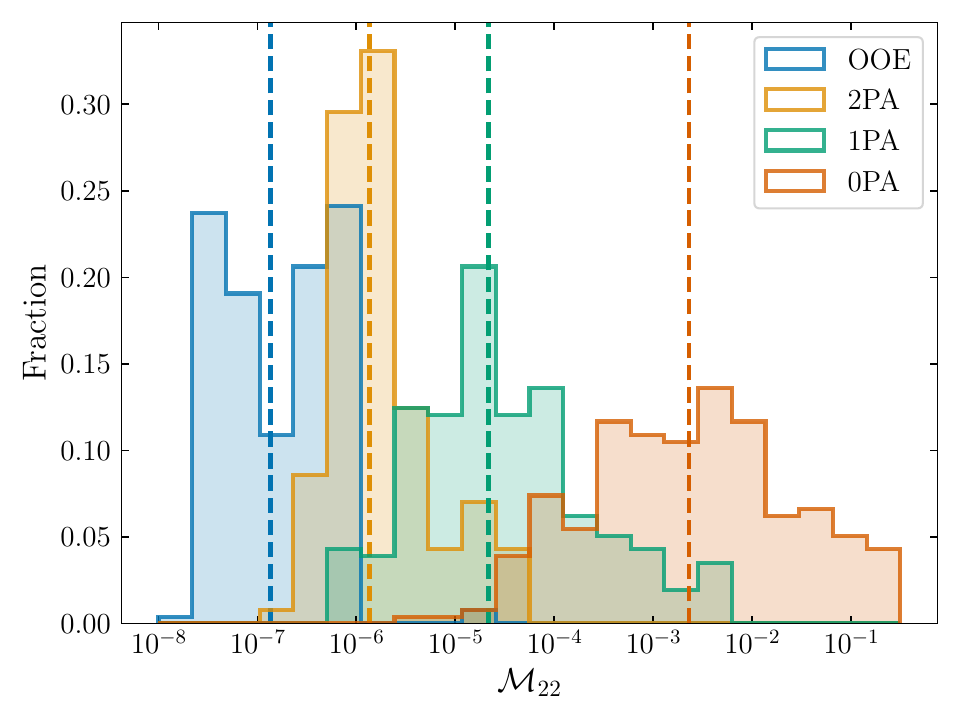} %Was there a reason this was not using the full column width?
	\caption{Histogram of EOB-OOE/NIT mismatch at different postadiabatic orders for $M = 60 M_\odot$ over $q \in [1,96]$, $e \in [0.1, 0.85]$ starting with an orbit-averaged frequency $f_0 = 10$\,Hz. Mismatches use the O5 design sensitivity curve and a low-frequency cutoff of 10 Hz. Dashed vertical lines mark the median mismatch.
		\label{fig:Model_Mismatch_Histograms}
	}
\end{figure}

\begin{figure*}
	\includegraphics[width=.9\textwidth]{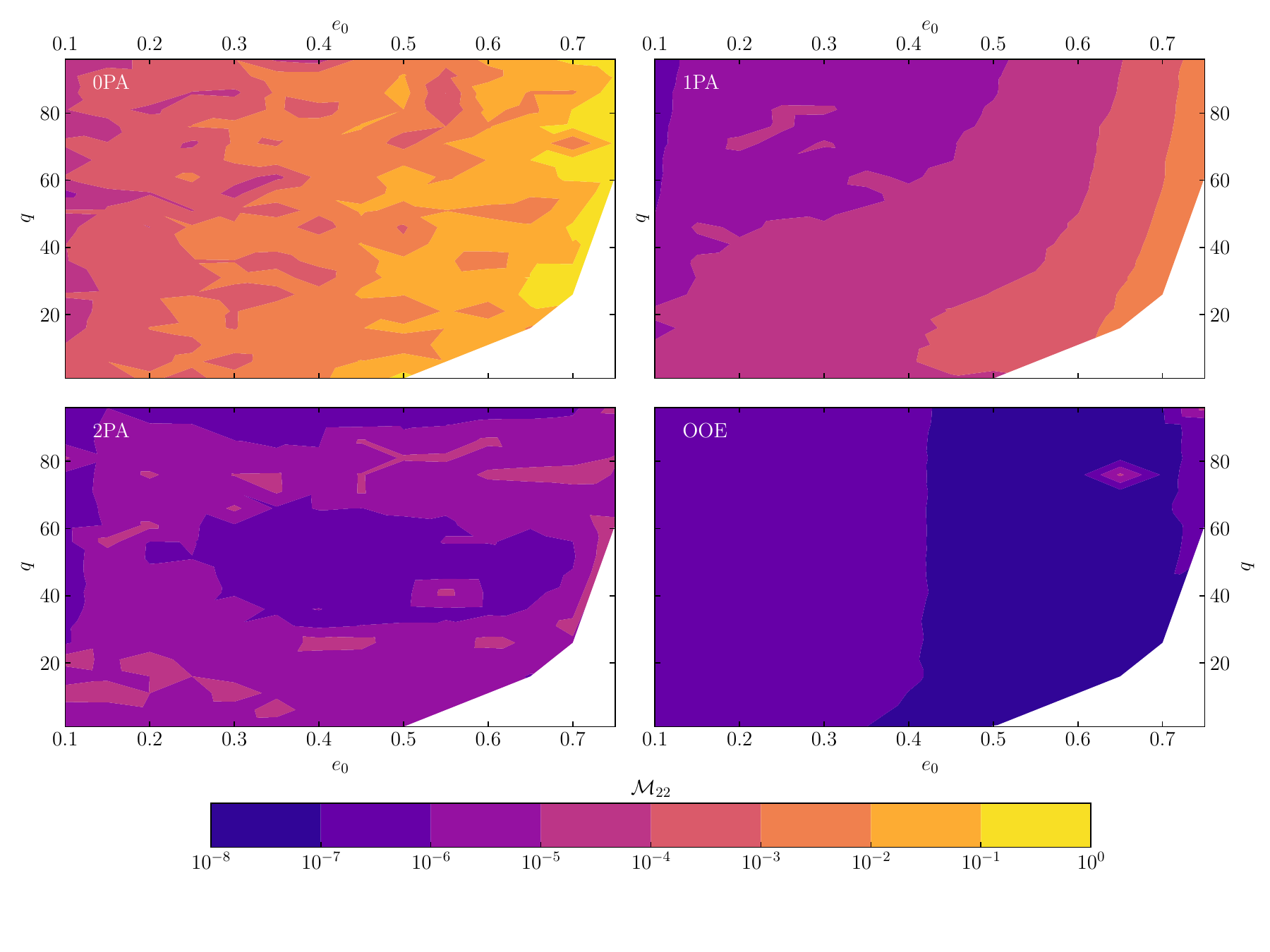}
	\caption{Contour plot of EOB-OOE/NIT mismatches at different postadiabatic orders for $M = 60 M_\odot$ over $q \in [1,96]$, $e \in [0.1, 0.85]$ starting with an orbit-averaged frequency $f_0 = 10$\,Hz. Mismatches use the O5 design sensitivity curve and a low-frequency cutoff of 10 Hz. White regions are systems that begin after the NIT handover conditions are violated.
		\label{fig:Model_Mismatch_Contours}
	}
\end{figure*}

We now examine the accuracy of different postadiabatic orders across parameter space. We fix the total mass at $60M_\odot$. We grid $q \in [1,96]$ in steps of 5 and $e \in [0.1, 0.85]$ in steps of 0.05 for 320 points in total. Each system is initialized at periastron (i.e., $\zeta_0 = 0$) with an orbit-averaged frequency of $f_0 = 10 \text{Hz}$. We compute $\mathcal{M}_{22}$ between EOB and the NIT/OOE waveforms using the O5 design sensitivity curve as outlined in Sec.~\ref{section:MismatchCalculation}.

Note that a small subset of these initial conditions can yield very short inspirals with low adiabaticity $\mathcal{Q} < 100$. Such systems start after the handover from NIT to EOB dynamics (Sec.~\ref{section:Handover}). This is especially true for high-eccentricity systems with $e_0>0.75$, suggesting that this method could be ill suited for such high eccentricities \cite{Fumagalli:2025rhc}. We therefore exclude systems that immediately transition to the EOB dynamics, since their waveforms would be identical by construction.

The results of this comparison are shown in Fig.~\ref{fig:Model_Mismatch_Histograms}. Adding more postadiabatic information significantly improves waveform accuracy. The 0PA waveforms have a median mismatch of $\sim 2 \times 10^{-3}$ with some becoming as large as $\sim 0.5$. The 1PA waveforms improve on this, with a median $2 \times 10^{-5}$ and a tail extending to $6 \times 10^{-3}$. The 2PA waveforms reach a median of $\mathcal{M}_{22} \sim 5 \times 10^{-6}$ and a maximum of $5 \times 10^{-5}$. Thus for comparable-mass systems, at least 2PA accuracy is required.

Figure~\ref{fig:Model_Mismatch_Contours} shows where these mismatches occur in parameter space. The 0PA and 1PA waveforms have the highest mismatches for comparable-mass ratio and high-eccentricity systems. These systems are the least adiabatic, where the NIT expansions are expected to perform worst. 1PA may suffice even for comparable-mass systems at low eccentricities, but more information is needed for higher eccentricities.

The 2PA waveforms have low mismatches throughout, with the highest mismatches appearing sporadically across the parameter space. These errors most likely stem from the transformations back to the EOB coordinates used for the plunge, which depend on $\zeta$ as well as $(\nu,p,e)$. This could be remedied by a more optimal choice of handover conditions than those used in Sec.~\ref{section:Handover}. However, these mismatches remain very low, so this does not significantly affect waveform accuracy.

Finally, the OOE waveforms have low mismatches across the entire parameter space, though they are slightly larger for smaller eccentricities, with seemingly no dependence on the mass ratio. Since these are generally smaller than the 2PA mismatches, the errors arising from the low-eccentricity limit of this choice of orbital element parametrization, interpolation, the ODE solver, and our method for sampling the waveform are all subdominant to the 2PA truncation error.

\subsection{Second-postadiabatic-order accuracy across parameter space} \label{section:2PAComparison}

We now examine the 2PA NIT accuracy across parameter space more thoroughly. We explore slices of the parameter space avoiding waveforms that are too long to generate (e.g., very low-mass, high-mass-ratio systems).
For each fixed total mass, we start at periastron with orbit-averaged frequency of $f_0 = 10$Hz and sample equally in eccentricity from $e \in [0.1,0.85]$ in steps of $0.05$. We sample in the mass ratio $q$ with steps $\Delta q$ as follows:
\begin{itemize}
	\item $M = 100 M_\odot$, $q \in [1,96]$, $\Delta q = 5$,
	\item $M = 50 M_\odot$, $q \in [1,96]$, $\Delta q = 5$,
	\item $M = 20 M_\odot$, $q \in [1,20]$, $\Delta q = 1$,
	\item $M = 10 M_\odot$, $q \in [1,10]$, $\Delta q = 0.5$.
\end{itemize}
We then compute $\mathcal{M}_{22}$ between EOB and the 2PA NIT waveforms using the O5 design sensitivity curve as in Sec.~\ref{section:MismatchCalculation}.
Again a small subset of these initial conditions (especially higher-mass, high-eccentricity, and low-mass-ratio systems) can yield very short inspirals that start after the NIT handover conditions, and thus, are excluded from the analysis. A more rigorous analysis of the NITs at eccentricities $e \gtrsim 0.85$ is left to future work, as the NIT method may be ill suited for such high eccentricities due to the nonadiabaticity of such systems and the need for a higher number of Fourier modes~\cite{Fumagalli:2025rhc,McCart:2021upc}.

\begin{figure}[tbp]
\includegraphics[width=1.0\columnwidth]{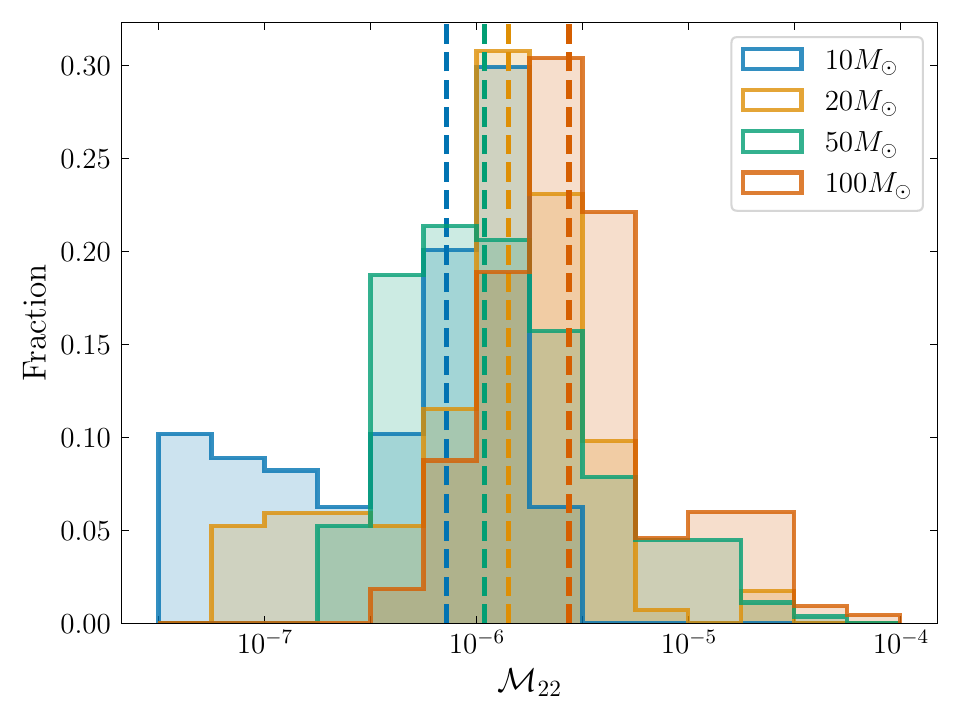}
\caption{Histogram of the EOB-2PA NIT mismatch for four total masses. Mismatches use the O5 design sensitivity curve and a low-frequency cutoff of 10 Hz. Dashed vertical lines mark the median mismatch for each total mass. 
\label{fig:2PA_Mismatch_Histograms}
}
\end{figure}

The results are shown in Fig.~\ref{fig:2PA_Mismatch_Histograms}. The 2PA NIT waveforms perform well across the entire parameter space, with median mismatch of $\mathcal{M}_{22} = 1.25 \times 10^{-6}$ and a maximum $\mathcal{M}_{22} = 8.05 \times 10^{-5}$. The lowest mismatches occur for low-mass systems with long inspirals, showing that the 2PA dynamics track the phase evolution for long-lived signals accurately. The worst mismatches occur for shorter-lived, high-mass signals, indicating that most of the error is accumulated in the late inspiral and transition to plunge, rather than the early inspiral. 

A further point is that, at fixed physical frequency (here, 10Hz), increasing the total mass raises the \textit{geometric} frequency: a system with, e.g., $e = 0.5$ is more relativistic at $100 M_\odot$ than at $10 M_\odot$. The $M = 100 M_\odot$ histogram in Fig.~\ref{fig:2PA_Mismatch_Histograms} thus samples much more relativistic systems than the $M = 10 M_\odot$ one, which may also contribute to the loss of accuracy. 

\begin{figure*}[tbp]
\includegraphics[width=0.9\textwidth]{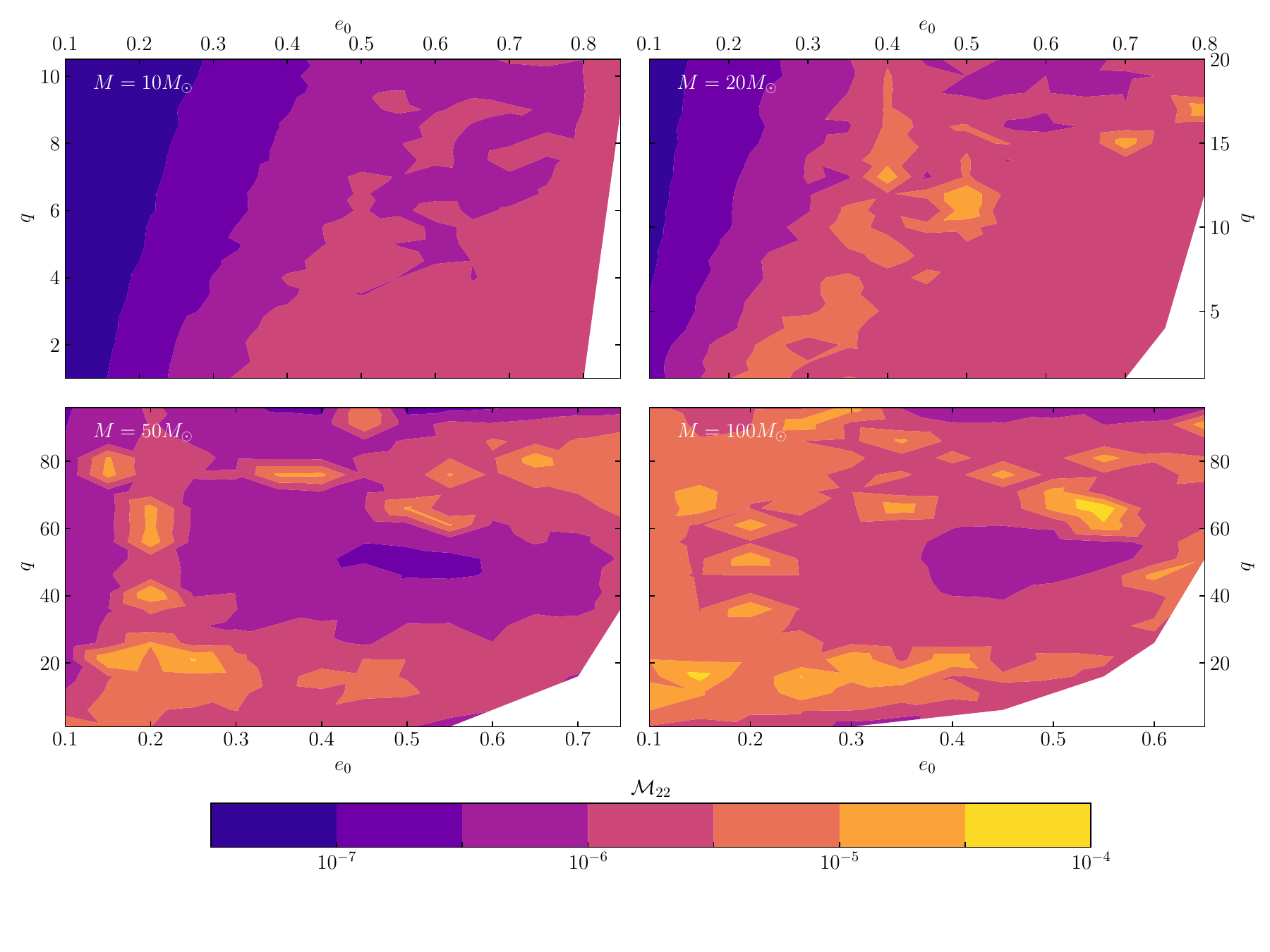}
\caption{Contour plot of the EOB-2PA NIT mismatch for four total masses. Mismatches use the O5 design sensitivity curve and a low-frequency cutoff of 10 Hz. White regions are systems that begin after the NIT handover conditions are violated.
\label{fig:2PA_Mismatch_Contours}
}
\end{figure*}

Figure~\ref{fig:2PA_Mismatch_Contours} shows contour plots of the 2PA NIT mismatches for the four total masses. For low-mass systems, the highest mismatches occur at larger eccentricities and smaller mass ratios, where the inspirals are less adiabatic. 
For $M = 10M_\odot$ in particular, the lowest mismatches are due to the  $\tilde{e} \leq 0.035$ condition triggering in the weak field and
transitioning to EOB dynamics relatively early. This along with our uniform sampling in parameter space results in the bimodality seen in Fig.~\ref{fig:2PA_Mismatch_Histograms}. For higher-mass systems, the highest mismatches occur sporadically across parameter space, likely due to our choice of handover conditions (Sec.~\ref{section:Handover}); a better choice could remedy this. To mitigate the errors from the $\zeta$-dependent transformation terms in the handover to EOB variables for the plunge, we produce the same plots for the inspiral section only in Appendix \ref{section:InspiralMismatches}.

Finally, the mismatch depends weakly on the mass ratio, degrading as the mass ratio decreases. Since even the $q = 1$ mismatches remain $< 10^{-5}$, this confirms that the 2PA NIT waveforms are accurate enough for all mass ratios.

\subsection{Speedup} \label{section:Runtime}
Now we explore the speedup achieved by using NIT dynamics compared to EOB dynamics across parameter space. We report the timings for the waveforms of the 2PA accuracy comparison in Sec.~\ref{section:PAComparison} and take the ratio of the EOB to NIT walltime. 

% Speedup histograms for the dynamics
\begin{figure}[tbp]
\includegraphics[width=0.95\columnwidth]{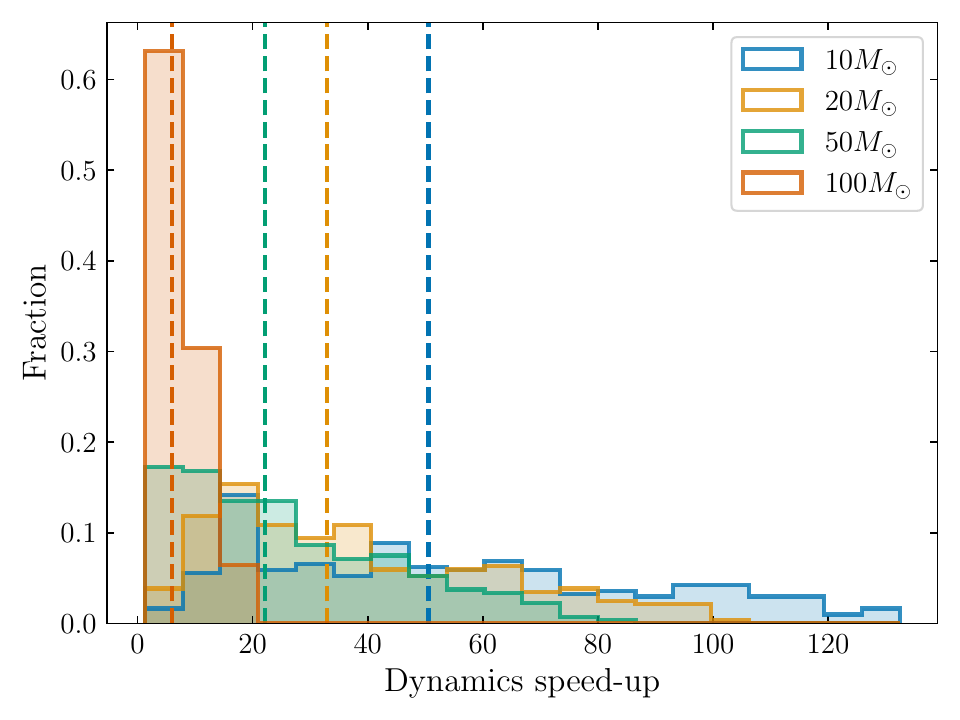}
\caption{Speed-up of the dynamics from using NIT rather than EOB dynamics for the parameters in Sec.~\ref{section:PAComparison}. The dashed vertical lines mark the median speedup for each total mass.
\label{fig:Dynamics_Speedup_Histograms}
}
\end{figure}

In Fig.~\ref{fig:Dynamics_Speedup_Histograms}, we show the speedup of producing the dynamics alone (including the changeover to the EOB transition to plunge) relative to the EOB dynamics for the parameters of Sec.~\ref{section:PAComparison}. Longer-lived, low-mass inspirals show the largest speedup, since they are dominated by the inspiral. For systems with $M = 10M_\odot$, the speedup for the dynamics ranges from $6.6$ to $132\times$, with a median of $\sim 50 \times$. The effect is less dramatic for heavier, shorter-lived systems where the dynamics cost is dominated by the transition to plunge. The dynamics speedup ranges from $1.25$ to $16.7\times$ with a median of  $\sim 6\times$ at $100M_\odot$. 

% Speedup histograms for the total waveform generation
\begin{figure}[tbp]
\includegraphics[width=0.95\columnwidth]{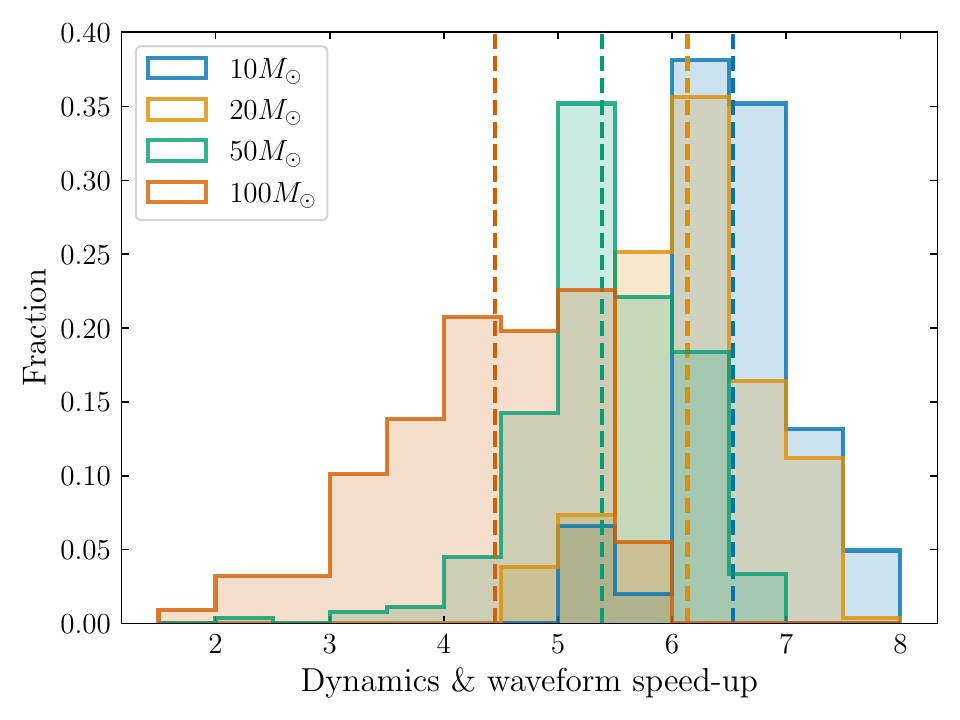}
\caption{Total speedup of the dynamics and waveform generation from using NIT rather than EOB dynamics for the parameters of Sec.~\ref{section:PAComparison}. The dashed vertical lines indicate the median speedup for each total mass.
\label{fig:Total_Speedup_Histograms}
}
\end{figure}

In Fig.~\ref{fig:Total_Speedup_Histograms} we show the total walltime speedup (dynamics plus waveform generation) of the NIT over the EOB approach, for the same parameters.
The total speedup depends mostly on total mass and mass ratio, ranging from $ 1.8$ to $ 7.8 \times$ with a median of $\sim 6 \times$ over the parameter space explored. 
It is lower than the speedup of the dynamics alone because waveform generation, not the dynamics, is now the computational bottleneck. 
This can be alleviated by interpolating the waveform modes as functions of the NIT variables, which we will implement in a follow-up paper \cite{Lynch2026SmallEccentricities}.
% Speedup as a function of total Mass and eccentrcity
\begin{figure}[tbp]
\includegraphics[width=0.9\columnwidth]{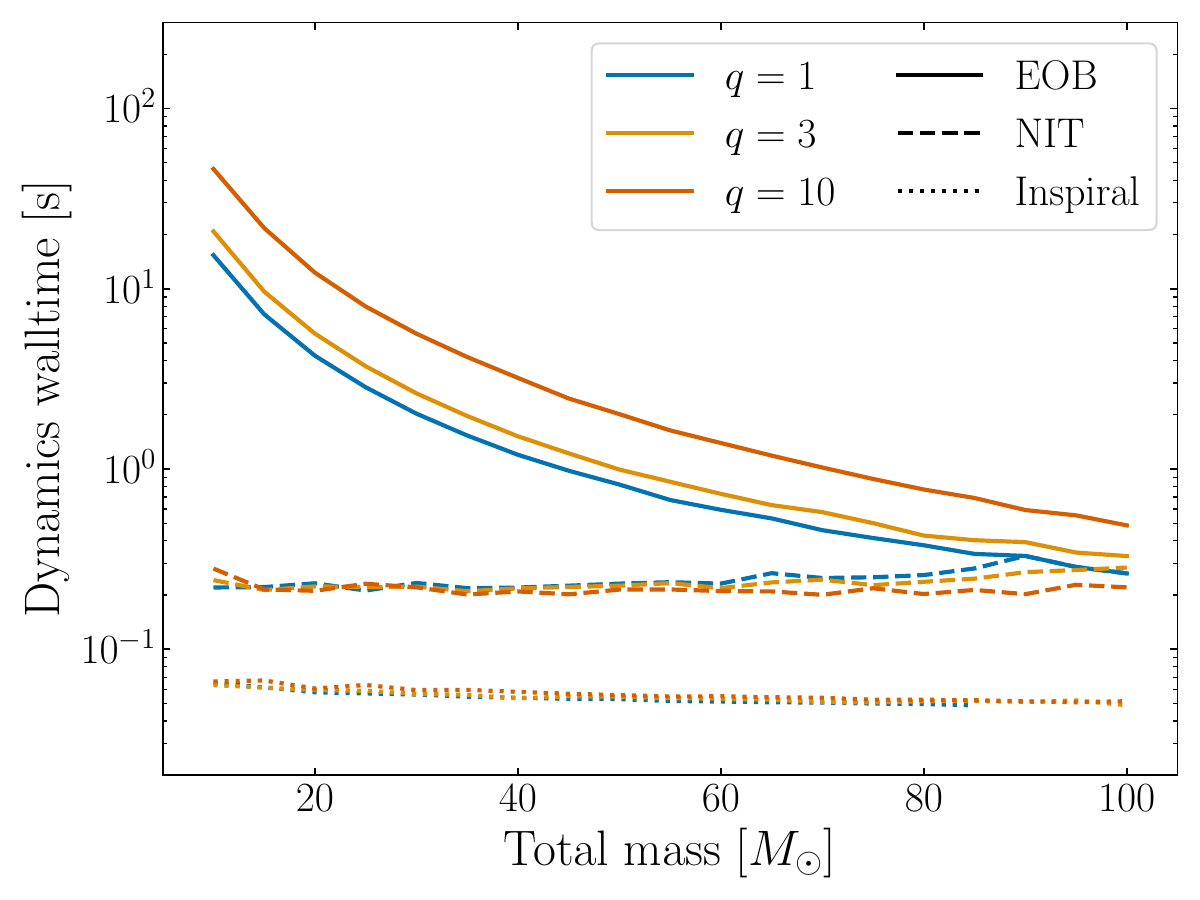}
\caption{Dynamics walltime versus total mass for a system starting at periastron, with orbit-averaged initial frequency of $10$Hz and initial eccentricity of $e_0 = 0.4$. We show the walltime of the EOB dynamics (blue), the NIT dynamics with the EOB transition to plunge (red), and NIT dynamics without the EOB transition to plunge (orange).
\label{fig:Dynamics_Walltime_vs_Mass}
}
\end{figure}

We now study how the speedup varies with total mass by examining the dynamics-only and total walltimes for a system starting at periastron with initial orbit-averaged frequency of $10$Hz and eccentricity of $e_0 = 0.4$ as total mass is varied. In Fig.~\ref{fig:Dynamics_Walltime_vs_Mass}, the dynamics cost is almost independent of total mass and most of it comes from the final transition to plunge which uses the EOB dynamics. 

\begin{figure}[tbp]
\includegraphics[width=0.9\columnwidth]{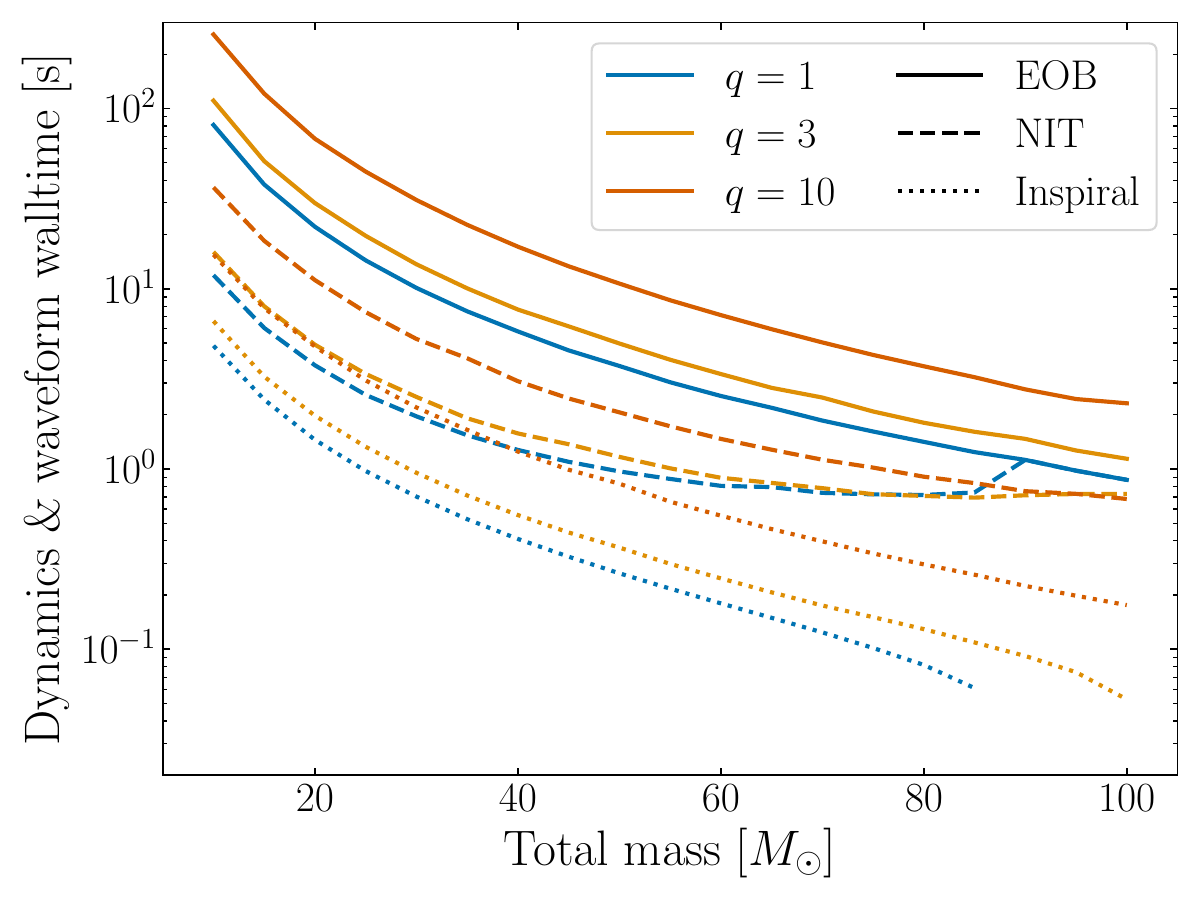}
\caption{The total walltime (dynamics and waveform generation) versus total mass for a system starting at periastron, with orbit-averaged initial frequency of $10$Hz and initial eccentricity of $e_0 = 0.4$. We show the walltime of the EOB dynamics (blue), the NIT dynamics with the EOB transition to plunge (red), and NIT dynamics without the EOB transition to plunge (orange).
\label{fig:Total_Walltime_vs_Mass}
}
\end{figure}

In Fig.~\ref{fig:Total_Walltime_vs_Mass}, the total walltime does increase with total mass, as expected since the waveform-generation cost scales with the length of the signal, which grows inversely with total mass. Again, most of the NIT cost comes from the final EOB transition to plunge. Despite this, using the NIT dynamics for the inspiral still yields a significant speedup of around $1.5 - 7 \times$, with a median of $\sim 4.5$ across the total masses examined.

% Speedup as a function of mass ratio
\begin{figure}[tbp]
\includegraphics[width=0.95\columnwidth]{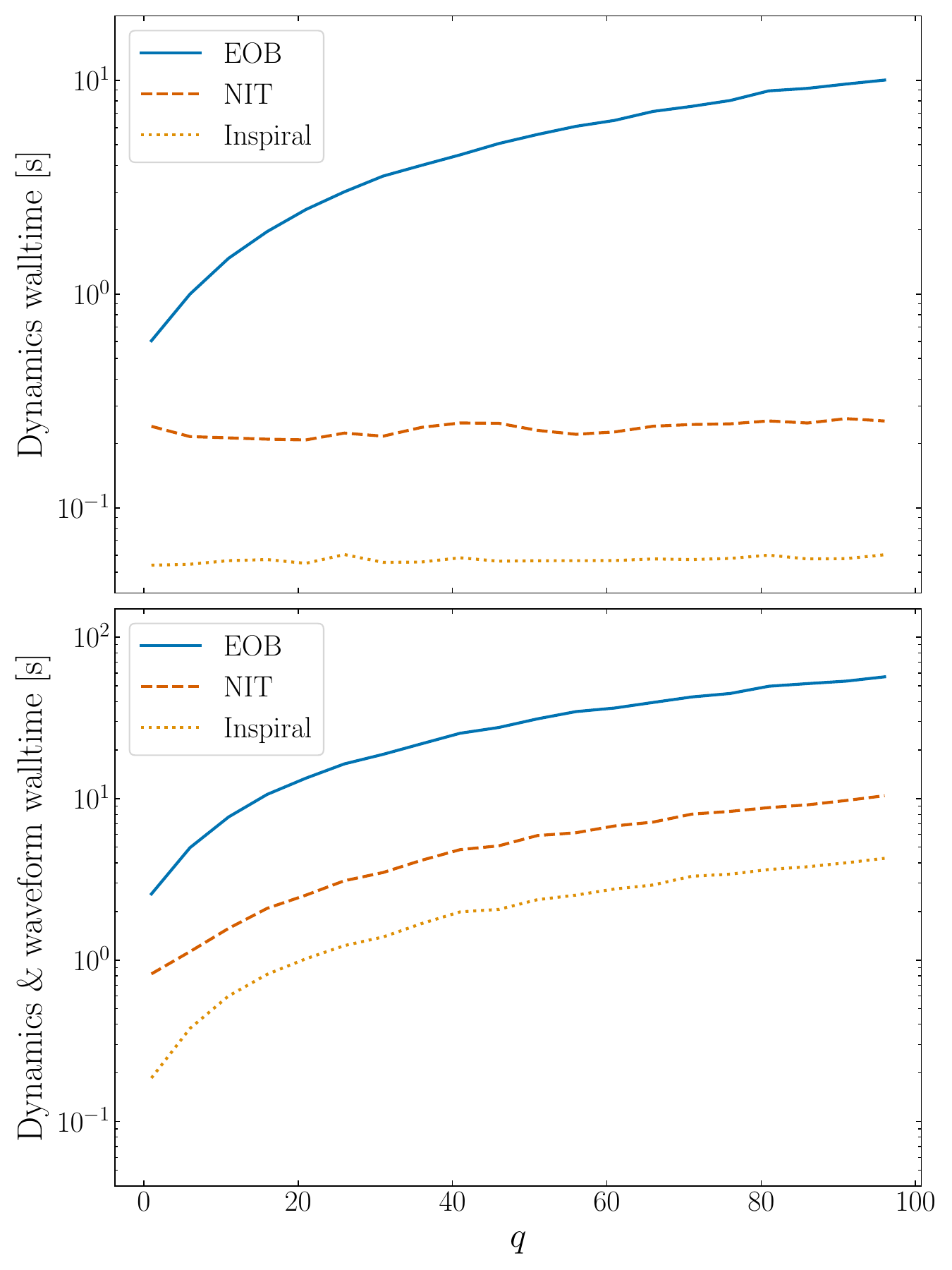}
\caption{
The walltime vs mass ratio for a system with $M = 60M_\odot$ started at periastron with initial orbit-averaged frequency of $10$Hz and initial eccentricity of $e_0 = 0.4$. The top panel shows the dynamics walltime; the bottom panel shows the total walltime of the dynamics and waveform generation. The blue curve denotes using the EOB dynamics, the red curve denotes using the NIT dynamics along with the EOB transition to plunge, and the orange curve denotes the time taken for only the NIT inspiral dynamics.
\label{fig:Walltime_vs_MassRatio} 
}
\end{figure}

Finally, we study how the speedup varies with mass ratio for a system $M = 60M_\odot$, started at periastron with initial orbit-averaged frequency of $10$Hz and initial eccentricity of $e_0 = 0.4$. In the top panel of Fig.~\ref{fig:Walltime_vs_MassRatio}, the EOB dynamics cost increases with mass ratio, while the NIT dynamics cost is independent of it.
In the bottom panel, the total walltime of both the EOB and NIT waveforms increases with mass ratio. Again, this is because waveform generation, now the bottleneck, scales with signal length, which grows with mass ratio. Nonetheless, the NIT inspiral still gives a significant $3 - 5.6\times$ speedup with a median of $\sim 5.3\times$ across the mass ratios examined.
In both panels, we still see that the NIT timings are dominated by the final EOB transition to plunge.

\section{Discussion} \label{section:Discussion}

% Summarise the method
In this paper, we tackled the cost of eccentric EOB inspiral dynamics by recasting the equations of motion into quasi-Keplerian osculating orbital elements and then applying near-identity (averaging) transformations to obtain a new set of equations of motion that are independent of the orbital phases. These can be solved efficiently on the radiation-reaction timescale, decoupling the computational cost and the mass ratio. We combined this with an efficient waveform sampling method which further reduces computational cost. Finally, we transformed back to the original EOB equations of motion to model the final transition to plunge. Comparing against waveforms generated in the standard way, we found that the NIT dynamics must be computed to at least second postadiabatic (2PA) order to accurately model comparable mass systems. This yields mismatches $\mathcal{M}_{22} \leq 8.05 \times 10^{-5}$ across the parameter space while reducing the dynamics cost by up to 2 orders of magnitude and achieving an overall waveform-generation speedup of between $\sim 1.5$ and $8 \times$.

% Discuss improvements that could be made to the method such as including higher postadiabatic orders, improving the handover conditions, etc.
Several improvements remain possible. While the 2PA NIT waveforms are sufficiently accurate for current and future ground-based detectors, the OOE waveforms suggest that higher postadiabatic orders would achieve even greater accuracy, which might be necessary for the highest-SNR MBHB signals one may see with LISA. Additionally, we find that the highest mismatches occur sporadically across parameter space, most likely due to our transformation back to the EOB coordinates for the final transition to plunge. Moreover, most of the computational cost is due to this same region which uses the slower EOB dynamics. Both points warrant a more detailed study of optimal handover conditions. A more physically motivated eccentric transition to plunge could further improve the accuracy and speed of the resulting waveforms \cite{Becker:2024xdi,Lhost:2024jmw}.

Our current parametrization of the dynamics is ill suited to the quasicircular limit.
A follow-up paper will use a different set of osculating orbital elements that are better suited to this limit. For systems with moderate eccentricities that evolve to be less than our cutoff eccentricity of $\tilde{e} < 0.035$, we will start the inspiral with the present parametrization and then transition to the new one for the remainder of the inspiral \cite{Lynch2026SmallEccentricities}.

In this work only eccentricities $e \leq 0.9$ were considered, as this was the arbitrarily chosen cutoff for our interpolation grid, and most of the systems with eccentricities above $e_{10 Hz} \gtrsim 0.7$ had initial conditions which violated our $\mathcal{Q} \leqslant 100$ handover condition. Methods that rely on orbit-averaging would be ill suited for very high eccentricities \cite{McCart:2021upc,Fumagalli:2025rhc}, and Fourier decomposing with respect to the orbital phase also scales poorly with eccentricity. A more detailed study where one does not rely on Fourier decomposition and interpolates to much higher eccentricities should be carried out in future work to understand the exact limitations of NITs in this limit.

For long signals, we are now dominated by the waveform-generation cost. Combining the averaged dynamics with a transformation to monotonically evolving phases \cite{Lynch:2024hco} would allow one to use a multivoice decomposition of the waveform modes \cite{Hughes:2021exa}. This can be trivially parallelized on both CPUs and GPUs to rapidly construct long inspiral signals as demonstrated by the \texttt{FastEMRIWaveforms}  package \cite{Chua:2020stf,Katz:2021yft,Speri:2023jte,Chapman-Bird:2025xtd}. Adapting this method to the variables presented here will also be explored in a follow-up paper \cite{Lynch2026SmallEccentricities}.

Another approach would be to reduce the need for dense time-domain sampling to produce faithful frequency-domain waveforms. Many ongoing works explore this using nonuniform time sampling, frequency-domain transformations via the stationary phase approximation \cite{Haberland:2026xvj} or shifted uniform asymptotics \cite{Morras:2025nlp,Klein:2014bua}, or transforming the waveform to time-frequency-domain \cite{Cornish:2020odn,Bandopadhyay:2025fyx}. All of these could be applied to our averaged dynamics to further reduce waveform-generation cost.

So far we have implemented only nonspinning dynamics and waveforms. Adding (anti)aligned spin is a straightforward extension. The only difficulty is interpolating over the values of the primary and secondary spins, but efficient multidimensional interpolation methods are well established and should suffice \cite{christian_chapman_bird_2025_15231132,Khalvati:2025znb}. Adding spin precession is less trivial, but still possible, requiring transformations that average over both eccentric and spin precession oscillations \cite{Lynch:2024hco,Drummond:2023wqc}.

The averaged dynamics trivially relate to the eccentric dynamics for EMRIs, modeled with the gravitational self-force formalism. Many efforts have already used information from the test-mass limit to improve EOB waveforms at large mass ratios \cite{Albertini:2022dmc,Albertini:2023aol,Albertini:2024rrs,Albertini:2024agg,Faggioli:2025hff,Nishimura:2026nse,Faggioli:2026alx}. Our framework can offer new calibration methods. For example, calibrating the test-mass limit of eccentric EOB radiation reaction force to eccentric self-force flux calculations \cite{Faggioli:2024ugn} would greatly reduce the dominant error of EOB waveforms for extreme mass ratios stemming from missing adiabatic order information \cite{Khalvati:2025znb}. Conversely, one could supplement adiabatic self-force models with postadiabatic information from eccentric EOB models to produce eccentric hybrid models, similar to what has been done with post-Newtonian information \cite{Burke:2023lno,Honet:2025gge,Trestini:2026tky}. 

Finally, the method presented here is agnostic to the specific EOB model, so long as the equations of motion take the form of Eqs.~\eqref{eq:EOB_EOM}. Having demonstrated it on a toy, inspiral-only EOB model, a natural next step would be to apply it to state-of-the-art eccentric EOB models, such as \texttt{TEOBResumS-Dali} \cite{Chiaramello:2020ehz,Nagar:2020xsk, Nagar:2021gss,Nagar:2021xnh,Placidi:2021rkh,Albanesi:2021rby,Albanesi:2022ywx,Albanesi:2022xge,Albanesi:2023bgi,Placidi:2023ofj,Nagar:2024dzj,Nagar:2024oyk}, \texttt{SEOBNRE} \cite{Cao:2017ndf,Liu:2019jpg,Liu:2021pkr}, and \texttt{SEOBNRv6EHM} \cite{Gamboa:2026jht,Pompili:2026yxq}. This would enable efficient, accurate parameter estimation of eccentric BBHs for both current and future detectors.
While more work remains, this method stands as an important first step toward efficient eccentric inspirals in the EOB framework.

\begin{acknowledgments}
PL thanks Chris Kavanagh for helpful discussions at the early stages of the project and Angelica Albertini for finding typos in an earlier version of the manscipt.
This work was funded by the European Union (ERC grant GWSky/ 101167314). Views and opinions expressed are however those of the author(s) only and do not necessarily reflect those of the European Union or the European Research Council Executive Agency. Neither the European Union nor the granting authority can be held responsible for them. 
MvdM further acknowledges financial support by 
the VILLUM Foundation (Grant No. VIL37766),
 and the DNRF Chair program (Grant No. DNRF162) by the Danish National Research Foundation.
The Center of Gravity is a Center of Excellence funded by the Danish National Research Foundation under Grant No. 184.
\end{acknowledgments}

\appendix
%\begin{widetext}
\section{Radial momentum expansion coefficients} \label{section:prCoefficients}
To obtain an expression for $p_{r_*}$ in terms of the orbital elements, we start from the EOB effective Hamiltonian given by Eq.~\eqref{eq:Heff} and rearrange it into a power series in $p_{r_*}$:
\begin{equation}
\begin{aligned}
 \hat{p}_{r*}^2 & = H_{\text{eff}}^2 - A \left(1 + \frac{p_\phi^2}{r^2} \right) \\& = p_{r_*}^2 + A Q_4  p_{r_*}^4 + A Q_6 p_{r_*}^6  + A Q_8 p_{r_*}^8 + \mathcal{O}(p_{r_*}^{10}),
\end{aligned}
\end{equation}
where $\hat{p}_{r*} = p_{r_*}$ when $Q = 0$, and the potentials $Q_4$, $Q_6$, and $Q_8$ are given explicitly in Appendix A of Ref.~\cite{Pompili:2023tna}. The quantity $\hat{p}_{r*}$ is defined entirely in terms of quantities with explicit expressions in $(p,e,\zeta)$. We can thus perturbatively invert this expansion to obtain $p_{r_*}$ as an approximate power series in $\hat{p}_{r*}$:
\begin{equation}
	p_{r_*}(\nu,p,e,\zeta) \simeq \text{sgn}(\sin(\zeta))  \sum_{n=1}^{7} c_n(\nu,p,e,\zeta) \hat{p}_{r*}^n(\nu,p,e,\zeta),
\end{equation}
where the nonzero coefficients $c_n$ are given by
\begin{subequations}
\begin{align}
	c_1 & = 1, \\
	c_3 & = -\frac{1}{2} A Q_4, \\
	c_5 & = \frac{1}{8} \left( 7 A^2 Q_4^2 - 4 A Q_6\right) \\
	c_7 & = \frac{1}{16} \left( -33 A^3 Q_4^3 + 36 A^2 Q_4 Q_6 - 8 A Q_8 \right).
\end{align}
\end{subequations}

\section{Fourier coefficients of functions via their reciprocal} \label{section:ReciprocalFourierCoefficients}

\begin{figure}[tbp]
\includegraphics[width=0.95\columnwidth]{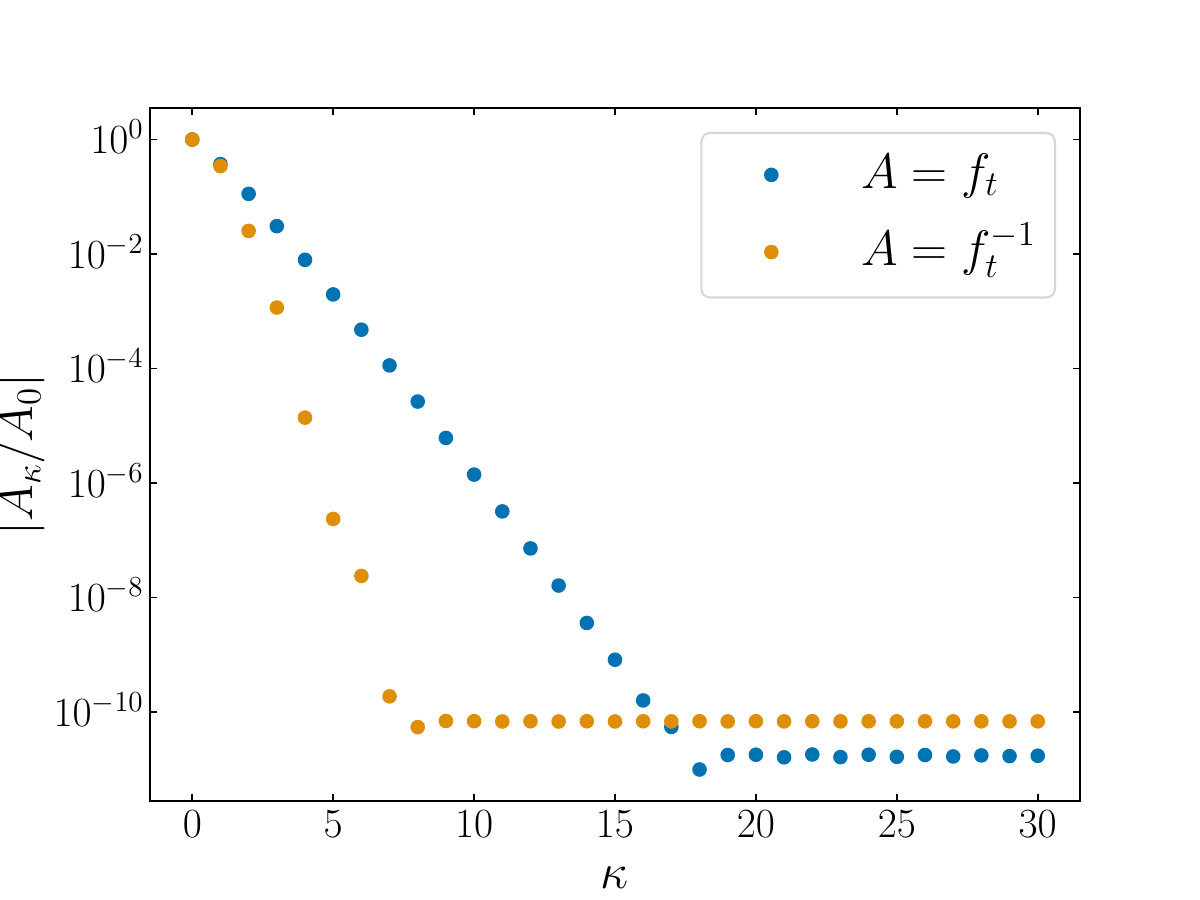}
\caption{Absolute values of the Fourier coefficients of $f_t(\zeta)$ (blue) and its reciprocal $f_t(\zeta)^{-1}$ (orange) normalized by the zeroth coefficient for $(q,p_0,e_0) = (1,20,0.4)$. The reciprocal $f_t(\zeta)^{-1}$ reaches its numerical floor much sooner than $f_t(\zeta)$, so it is well approximated by a low-order Fourier expansion.
\label{fig:ft_and_Reciprocal}
}
\end{figure}

Here we summarize a method to compute the Fourier coefficients of a function $A(\zeta)$ given those of its reciprocal $A(\zeta)^{-1}$. This is particularly useful for the functions $f_t(\zeta)$ and $f_\phi(\zeta)$, which appear Eqs.~\eqref{eq:OOE_EoM}. In Fig.~\ref{fig:ft_and_Reciprocal}, we plot the absolute values of the Fourier coefficients of $f_t(\zeta)$ and its reciprocal $f_t(\zeta)^{-1}$ (normalized by the zeroth coefficient) with the parameters $(q,p_0,e_0) = (1,20,0.4)$. We see that even at moderate eccentricities, $f_t(\zeta)$ would not be well approximated by a low-order Fourier expansion, while $f_t(\zeta)^{-1}$ would. We now establish a method to exploit this observation.

Let the Fourier coefficients of a function $A(\zeta)$ be denoted $A_\kappa$ as in Eq.~\eqref{eq:Fourier}.
Let $B(\zeta)$ be the reciprocal function $A(\zeta)^{-1}$ with Fourier coefficients $B_{\kappa'}$.
Assuming $A(\zeta) \neq 0$, so that $A(\zeta) B(\zeta) = 1$, we can write:
\begin{equation}
	\sum_{\kappa} A_\kappa e^{i \kappa \zeta} \sum_{\kappa'} B_{\kappa'} e^{i \kappa' \zeta}	  =  \sum_\kappa  \sum_{\kappa'} A_{\kappa'} B_{\kappa - \kappa'}  e^{i \kappa  \zeta} =  1.
\end{equation}
Since the right-hand side is a constant that is independent of $\zeta$, this must mean that the above sum is $1$ for $\kappa = 0$ and otherwise zero. Thus we have the following relationship between the Fourier coefficients of $A$ and $B$:
\begin{equation}
	\sum_{\kappa'} A_{\kappa'} B_{\kappa - \kappa'} = \delta_{\kappa,0},
\end{equation}
where $\delta_{\kappa,0}$ is the Kronecker delta.

Note that we are restricted to a finite number of Fourier coefficients for both $A$ and $B$ such that $-\kappa_\text{max} \leq \kappa \leq \kappa_\text{max}$ and $-\kappa'_\text{max} \leq \kappa' \leq \kappa'_\text{max}$. We define the $N = 2\kappa_\text{max} + 1$ dimensional column vector of Fourier coefficients for $A$ as $\vec{A} = (A_{-\kappa_\text{max}}, \ldots, A_{\kappa_\text{max}})^T$. From the coefficients of $B_\kappa'$, we can construct the $N \times N$ Toeplitz matrix $\mathbf{G}$ as
\begin{equation}
	G_{i,j} = \begin{cases}
		 B_{i - j} & |i-j|	\leq \kappa_\text{max}' \\
		 0 & |i-j|	> \kappa_\text{max}'.
	\end{cases}
\end{equation}
Explicitly, the matrix $\mathbf{G}$ has the following form:
\begin{equation}
	\mathbf{G} = \begin{bmatrix}
	B_0 & B_{-1} & B_{-2} & \cdots \\
	B_1 & B_0 & B_{-1} & \cdots  \\
	B_2 & B_1 & B_0 & \cdots  \\
	\vdots & \vdots & \vdots & \ddots 
	\end{bmatrix}.
\end{equation}

This allows us to write the above relationship between the Fourier coefficients of $A$ and $B$ in matrix form as $ \mathbf{G} \vec{A}  = \vec{e}_0$,
where $\vec{e}_0$ is a column vector of length $N$ with a $1$ at the index corresponding to the $\kappa = 0$ entry and zeros everywhere else. Thus, so long as the matrix $\mathbf{G}$ is invertible, we can solve for the Fourier coefficients of $A$ given the Fourier coefficients of $B$ via $\vec{A} = \mathbf{G}^{-1} \vec{e}_0$.

\begin{figure}[tbp]
\includegraphics[width=0.95\columnwidth]{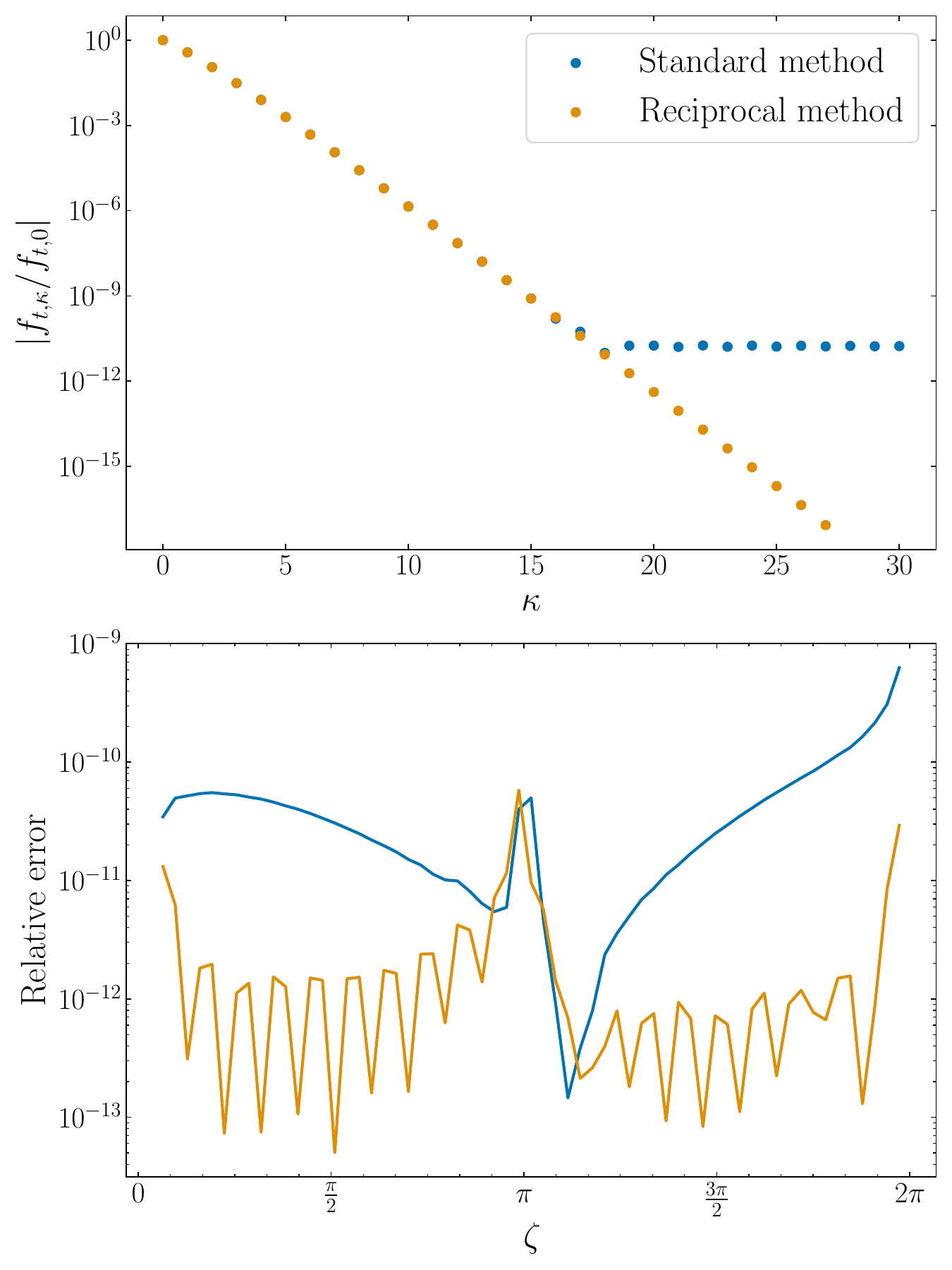}
\caption{Top panel: the Fourier coefficients of $f_t(\zeta)$ from standard method (blue) and the reciprocal method (orange). Bottom panel: the relative error between the original function and its Fourier series for  each method.
\label{fig:Reciprocal_Method_Convergence}
}
\end{figure}
 
To demonstrate this method, we first compute the Fourier coefficients of $f_t(\zeta)$ as outlined in Sec.~\ref{section:FourierDecomposition} using $\kappa_\text{max} = 30$. We then calculate the Fourier coefficients of $f_t(\zeta)^{-1}$ using $\kappa'_\text{max} = 10$, and then use the above method to compute the Fourier coefficients of $f_t(\zeta)$ for $-30 \le \kappa \le 30$. We show the results in Fig.~\ref{fig:Reciprocal_Method_Convergence}. We plot the absolute value of the modes $|f_{t,\kappa} |$ in the top panel. We see that the standard method reaches a numerical floor of $\sim 10^{-9}$ at $\kappa = \pm 18$. The reciprocal method, on the other hand, is able to compute Fourier coefficients up a numerical floor of $\sim 10^{-15}$ which occurs at $\kappa = \pm 28$, thus allowing us to resolve more of the harmonic content of $f_t(\zeta)$. This is confirmed in the bottom panel where we plot the relative error between the original function and its Fourier series computed with each method. The $L_2$ norm of the relative error is $8.2 \times 10^{-10}$ for the standard method and  $7 \times 10^{-11}$ for the reciprocal method. Thus, by using only $21$ Fourier coefficients of the reciprocal function, we are able to compute the $61$ Fourier coefficients of the original function to higher accuracy than the standard method.

\begin{widetext}
\section{Radiation-Reaction Force to 3PN order} \label{section:RRForce}
The azimuthal and radial components of the radiation-reaction force for nonspinning eccentric binaries to 3PN order can be expressed in terms of EOB coordinates as follows: 
\begin{align}
F_\phi = &
-\frac{\pi \nu}{225 r^6}
\left[
750 p_\phi^2
+ r \left( 5010 + 15390 p_r^2 r - 49 p_r^6 r^3 \right)
\right]
\nonumber\\[0.5ex]
&
+\frac{8 p_r p_\phi \nu^2}{1575 r^8}
\left[
31212 p_\phi^4
+ p_\phi^2 r \left( 17422 - 45261 p_r^2 r \right)
+ 2 r^2 \left( 12049 - 6160 p_r^2 r + 351 p_r^4 r^2 \right)
\right]
\nonumber\\[0.5ex]
&
+\frac{8 p_\phi \nu}{15 r^5}
\left[
10 p_\phi^2 - r \left( 22 + 29 p_r^2 r \right)
\right]
\nonumber\\[0.5ex]
&
+\frac{p_\phi \nu}{105 r^7}
\left[
-6 p_\phi^4 (93 + 4 \nu)
+ p_\phi^2 r \left( -3833 - 484 \nu + 12 p_r^2 r (30 + 83 \nu) \right)
\right.
\nonumber\\
&\qquad\qquad\left.
+ r^2 \left(
6213 + 1684 \nu
+ 12 p_r^4 r^2 (29 + 55 \nu)
+ p_r^2 r (15563 + 8324 \nu)
\right)
\right]
\nonumber\\[0.5ex]
&
+\frac{\pi \nu}{37800 r^8}
\left[
63000 p_\phi^4 (1 + \nu)
- 180 p_\phi^2 r \left( -24949 + 762 \nu + 700 p_r^2 r (1 + \nu) \right)
\right.
\notag\\
&\qquad\left.
+ r^2 \left(
-146 p_r^6 r^3 (-151 + 30 \nu)
+ 180 (-6817 + 24100 \nu)
+ 360 p_r^2 r (-2656 + 103379 \nu)
+ 15 p_r^4 r^2 (-267293 + 1382300 \nu)
\right)
\right]
\nonumber\\[0.5ex]
&
+\frac{p_\phi \nu}{5670 r^9}
\left[
18 p_\phi^6 (1058 + 2059 \nu + 258 \nu^2)
\right.
\nonumber\\
&\qquad\qquad
- 9 p_\phi^4 r \left(
-28869 - 13051 \nu + 16912 \nu^2
+ 6 p_r^2 r (-1021 + 2296 \nu + 2801 \nu^2)
\right)
\nonumber\\
&\qquad\qquad
- 3 p_\phi^2 r^2 \left(
-78397 - 88203 \nu - 93300 \nu^2
+ 36 p_r^4 r^2 (583 - 397 \nu + 750 \nu^2)
\right.
\nonumber\\
&\qquad\qquad\qquad\qquad\left.
+ 6 p_r^2 r (30166 + 47051 \nu + 16844 \nu^2)
\right)
\nonumber\\
&\qquad\qquad\left.
- r^3 \left(
18 p_r^6 r^3 (43 - 479 \nu + 1159 \nu^2)
+ 9 p_r^4 r^2 (21249 + 79681 \nu + 21128 \nu^2)
\right.\right.
\nonumber\\
&\qquad\qquad\qquad\qquad\left.\left.
+ 4 (-8294 + 174663 \nu + 37620 \nu^2)
+ 3 p_r^2 r (192199 + 1987011 \nu + 147540 \nu^2)
\right)
\right]
\nonumber\\[0.5ex]
&
+\frac{p_\phi \nu}{165375 r^8}
\left[
3 p_\phi^2 \left(
2697337 + 5812240 \gamma_E - 1901200 \pi^2 - 37330160 \log 2 + 49141890 \log 3
\right)
\right.
\nonumber\\
&\qquad\qquad
+ r \left(
25535157 - 89853333 p_r^2 r - 488318040 p_r^4 r^2 - 471411612 p_r^6 r^3
\right.
\nonumber\\
&\qquad\qquad\qquad\qquad
- 179760 \gamma_E (1 + 71 p_r^2 r)
+ 58800 \pi^2 (1 + 71 p_r^2 r)
+ 146504400 \log 2
\nonumber\\
&\qquad\qquad\qquad\qquad
+ 116664240 p_r^2 r \log 2
+ 14760453120 p_r^4 r^2 \log 2
- 156376435712 p_r^6 r^3 \log 2
\nonumber\\
&\qquad\qquad\qquad\qquad
- 147425670 \log 3
- 147425670 p_r^2 r \log 3
- 8845540200 p_r^4 r^2 \log 3
\nonumber\\
&\qquad\qquad\qquad\qquad\left.
+ 45510304329 p_r^6 r^3 \log 3
+ 36572265625 p_r^6 r^3 \log 5
\right)
\nonumber\\
&\qquad\qquad\left.
- 269640 \left( 97 p_\phi^2 - r (1 + 71 p_r^2 r) \right) \log r
\right]
\nonumber\\[0.5ex]
&
+\frac{p_\phi \nu}{6237000 r^{11}}
\left[
-450 p_\phi^8 (30034 + 96591 \nu + 137493 \nu^2 + 33708 \nu^3)
\right.
\nonumber\\
&\qquad\qquad
+ 225 p_\phi^6 r \left(
-980417 - 2276815 \nu - 784735 \nu^2 + 1633172 \nu^3
\right.
\nonumber\\
&\qquad\qquad\qquad\qquad\left.
+ 4 p_r^2 r (-94982 + 82077 \nu + 215352 \nu^2 + 205641 \nu^3)
\right)
\nonumber\\
&\qquad\qquad
+ 3 p_\phi^4 r^2 \left(
-781289023 + 25 (-43372039 + 767151 \pi^2)\nu + 722217825 \nu^2 - 247891500 \nu^3
\right.
\nonumber\\
&\qquad\qquad\qquad\qquad
+ 450 p_r^4 r^2 (-62971 + 140231 \nu + 442425 \nu^2 + 413998 \nu^3)
\nonumber\\
&\qquad\qquad\qquad\qquad\left.
+ 75 p_r^2 r (-1299933 + 10305311 \nu + 11856847 \nu^2 + 2297668 \nu^3)
\right)
\nonumber\\
&\qquad\qquad
+ p_\phi^2 r^3 \left(
900 p_r^6 r^3 (136450 - 149410 \nu - 275497 \nu^2 + 100135 \nu^3)
\right.
\nonumber\\
&\qquad\qquad\qquad\qquad
+ 225 p_r^4 r^2 (7472203 + 1277071 \nu + 8225939 \nu^2 + 2455644 \nu^3)
\nonumber\\
&\qquad\qquad\qquad\qquad
- 18 p_r^2 r \left(
-890506199 + 275 (-2919463 + 92988 \pi^2)\nu - 225199425 \nu^2 + 40233900 \nu^3
\right)
\nonumber\\
&\qquad\qquad\qquad\qquad\left.
+ 10 \left(
-284687488 + 15 (-11546167 + 1534302 \pi^2)\nu - 255988890 \nu^2 + 43794000 \nu^3
\right)
\right)
\nonumber\\
&\qquad\qquad
+ r^4 \left(
3 p_r^4 r^2 \left(
-989643623 + 25 (29305189 + 2045736 \pi^2)\nu + 1528454925 \nu^2 - 51998700 \nu^3
\right)
\right.
\nonumber\\
&\qquad\qquad\qquad\qquad
+ 900 p_r^8 r^4 (-10181 - 39439 \nu - 26926 \nu^2 + 15373 \nu^3)
\nonumber\\
&\qquad\qquad\qquad\qquad
+ 45 p_r^6 r^3 (-2474419 + 4183945 \nu + 14729025 \nu^2 + 3001420 \nu^3)
\nonumber\\
&\qquad\qquad\qquad\qquad
- 5 \left(
133552196 + 5 (-326935222 + 19008297 \pi^2)\nu - 84294180 \nu^2 + 10060560 \nu^3
\right)
\nonumber\\
&\qquad\qquad\qquad\qquad\left.
+ p_r^2 r \left(
-8758541524 - 225 (-156865186 + 4176711 \pi^2)\nu + 8786758500 \nu^2 + 304556400 \nu^3
\right)
\right)
\nonumber\\
&\qquad\qquad\left.
+ 6779520 r^2 \left(
36 p_\phi^4 + p_\phi^2 r (25 - 243 p_r^2 r) + r^2 (35 + 136 p_r^2 r + 36 p_r^4 r^2)
\right)\log r
\right],
\end{align}
\end{widetext}

\begin{widetext}
\begin{align}
F_r ={}&
-\frac{\pi p_r p_\phi \nu}{90 r^6}
\left(
2304 + 40 p_r^2 r + 5 p_r^4 r^2
\right)
\nonumber\\[0.5ex]
&\quad
+\frac{4 p_r \nu}{15 r^5}
\left[
-79 p_\phi^2 + r \left( 55 + 38 p_r^2 r \right)
\right]
\nonumber\\[0.5ex]
&\quad
-\frac{8 p_r^2 \nu^2}{1575 r^8}
\left[
-60867 p_\phi^4
+ 2 p_\phi^2 r \left( 35731 + 31563 p_r^2 r \right)
+ 2 r^2 \left( 7808 + 47500 p_r^2 r + 23409 p_r^4 r^2 \right)
\right]
\nonumber\\[0.5ex]
&\quad
+\frac{p_r \nu}{105 r^7}
\left[
-12 p_\phi^4 (30 + 83 \nu)
- p_\phi^2 r \left( 3317 + 3032 \nu + 6 p_r^2 r (-221 + 98 \nu) \right)
\right.
\nonumber\\
&\qquad\qquad\left.
+ r^2 \left(
13815 + 5204 \nu
+ 12 p_r^4 r^2 (93 + 4 \nu)
+ p_r^2 r (14057 + 5800 \nu)
\right)
\right]
\nonumber\\[0.5ex]
&\quad
+\frac{\pi p_r p_\phi \nu}{6300 r^8}
\left[
7 p_\phi^2 \left( -52800 - 200 p_r^2 r + 73 p_r^4 r^2 \right)(1 + \nu)
\right.
\nonumber\\
&\qquad\qquad\left.
+ r \left(
511 p_r^6 r^3 (1 + \nu)
+ 2 p_r^4 r^2 (-7951 + 1956 \nu)
+ 60 (22415 + 18004 \nu)
+ 5 p_r^2 r (256181 + 433724 \nu)
\right)
\right]
\nonumber\\[0.5ex]
&\quad
-\frac{p_r \nu}{5670 r^9}
\left[
-18 p_\phi^6 (-2005 + 8947 \nu + 8661 \nu^2)
\right.
\nonumber\\
&\qquad\qquad
- 9 p_\phi^4 r \left(
-18623 + 36921 \nu - 9230 \nu^2
+ 54 p_r^2 r (12 - 317 \nu + 138 \nu^2)
\right)
\nonumber\\
&\qquad\qquad
+ 3 p_\phi^2 r^2 \left(
678941 - 482793 \nu - 153972 \nu^2
+ 6 p_r^4 r^2 (5247 + 10774 \nu + 131 \nu^2)
\right.
\nonumber\\
&\qquad\qquad\qquad\qquad\left.
- 6 p_r^2 r (-84937 - 38240 \nu + 17458 \nu^2)
\right)
\nonumber\\
&\qquad\qquad\left.
+ r^3 \left(
36 p_r^6 r^3 (1058 + 2059 \nu + 258 \nu^2)
- 9 p_r^4 r^2 (-70131 - 54265 \nu + 17170 \nu^2)
\right.\right.
\nonumber\\
&\qquad\qquad\qquad\qquad\left.\left.
+ 3 p_r^2 r (-731279 + 1247585 \nu + 102884 \nu^2)
+ 4 (-996800 + 485649 \nu + 138222 \nu^2)
\right)
\right]
\nonumber\\[0.5ex]
&\quad
+\frac{p_r \nu}{330750 r^8}
\left[
p_\phi^2 \left(
62660418 - 113967840 \gamma_E + 37279200 \pi^2
+ 976636080 p_r^2 r + 942823224 p_r^4 r^2
\right.\right.
\nonumber\\
&\qquad\qquad\qquad\qquad
\left.\left.
+ 662595360 \log 2
- 29520906240 p_r^2 r \log 2
+ 312752871424 p_r^4 r^2 \log 2
- 884554020 \log 3
\right.\right.
\nonumber\\
&\qquad\qquad\qquad\qquad
\left.\left.
+ 17691080400 p_r^2 r \log 3
- 91020608658 p_r^4 r^2 \log 3
- 73144531250 p_r^4 r^2 \log 5
\right)
\right.
\nonumber\\
&\qquad\qquad
+ r \left(
4593918 - 1804655142 p_r^2 r - 3758925408 p_r^4 r^2
+ 359520 \gamma_E (413 + 703 p_r^2 r)
\right.
\nonumber\\
&\qquad\qquad\qquad\qquad
- 117600 \pi^2 (413 + 703 p_r^2 r)
- 593567520 \log 2
+ 58004117920 p_r^2 r \log 2
- 737028655104 p_r^4 r^2 \log 2
\nonumber\\
&\qquad\qquad\qquad\qquad\left.
+ 884554020 \log 3
- 33907904100 p_r^2 r \log 3
+ 200602109169 p_r^4 r^2 \log 3
+ 182861328125 p_r^4 r^2 \log 5
\right)
\nonumber\\
&\qquad\qquad\left.
+ 539280 \left( 317 p_\phi^2 - r (413 + 703 p_r^2 r) \right) \log r
\right]
\nonumber\\[0.5ex]
&\quad
+\frac{p_r \nu}{6237000 r^{11}}
\left[
-900 p_\phi^8 (-64948 + 178668 \nu + 352845 \nu^2 + 239349 \nu^3)
\right.
\nonumber\\
&\qquad\qquad
- 225 p_\phi^6 r \left(
-5246373 + 15930903 \nu + 10170659 \nu^2 - 2711680 \nu^3
\right.
\nonumber\\
&\qquad\qquad\qquad\qquad\left.
+ 2 p_r^2 r (-248981 + 227511 \nu + 1052289 \nu^2 + 1174578 \nu^3)
\right)
\nonumber\\
&\qquad\qquad
+ 3 p_\phi^4 r^2 \left(
-3075491833 + 25 (-13517149 + 4602906 \pi^2)\nu + 875931675 \nu^2 - 746247900 \nu^3
\right.
\nonumber\\
&\qquad\qquad\qquad\qquad
+ 300 p_r^4 r^2 (-16314 + 535774 \nu + 825469 \nu^2 + 34697 \nu^3)
\nonumber\\
&\qquad\qquad\qquad\qquad\left.
- 75 p_r^2 r (1736587 - 35658833 \nu - 11682157 \nu^2 + 2222400 \nu^3)
\right)
\nonumber\\
&\qquad\qquad
+ p_\phi^2 r^3 \left(
16924958492 - 119925 (-129238 + 2079 \pi^2)\nu - 5035306500 \nu^2 + 2368641600 \nu^3
\right.
\nonumber\\
&\qquad\qquad\qquad\qquad
+ 450 p_r^6 r^3 (230600 + 755015 \nu + 1016303 \nu^2 + 205210 \nu^3)
\nonumber\\
&\qquad\qquad\qquad\qquad
- 45 p_r^4 r^2 (-40018759 - 152321395 \nu - 71820895 \nu^2 + 38470240 \nu^3)
\nonumber\\
&\qquad\qquad\qquad\qquad\left.
- 3 p_r^2 r (-8508858414 + 25 (-345530410 + 7415793 \pi^2)\nu + 2840398950 \nu^2 + 411806200 \nu^3)
\right)
\nonumber\\
&\qquad\qquad
+ r^4 \left(
-313156972 + 25 (-967955966 + 5540535 \pi^2)\nu + 1448415900 \nu^2 - 528354000 \nu^3
\right.
\nonumber\\
&\qquad\qquad\qquad\qquad
+ 900 p_r^8 r^4 (30034 + 96591 \nu + 137493 \nu^2 + 33708 \nu^3)
\nonumber\\
&\qquad\qquad\qquad\qquad
- 225 p_r^6 r^3 (-1629369 - 8977187 \nu - 4638091 \nu^2 + 3066992 \nu^3)
\nonumber\\
&\qquad\qquad\qquad\qquad
- 3 p_r^4 r^2 (-2907765367 + 175 (-17557025 + 438372 \pi^2)\nu + 788417625 \nu^2 + 63942700 \nu^3)
\nonumber\\
&\qquad\qquad\qquad\qquad\left.
+ 2 p_r^2 r (-6499648016 - 75 (21154117 + 3665277 \pi^2)\nu + 1521061350 \nu^2 + 866592600 \nu^3)
\right)
\nonumber\\
&\qquad\qquad\left.
+ 6779520 r^2 \left(
171 p_\phi^4 - 4 p_\phi^2 r (77 + 72 p_r^2 r) + r^2 (233 + 223 p_r^2 r - 144 p_r^4 r^2)
\right)\log r
\right].
\end{align}
\end{widetext}

\begin{widetext}
\section{Waveform mode expressions to 3PN order} \label{section:WaveformModes} 
The $(\ell,m) = (2,2)$ spherical-harmonic mode amplitude for the waveform strain for eccentric nonspinning binaries to 3PN order can be expressed in terms of EOB coordinates as follows:
\begin{subequations}
\begin{align}
\mathcal{A}_{22} &=
-8\sqrt{\frac{\pi}{5}}\,\nu
\left(
\mathcal{A}_{22}^{\mathrm R}
+ i\,p_\phi\,\mathcal{A}_{22}^{\mathrm I}
\right),
\\
\alpha_\nu &\coloneqq -35 + 13\nu + 17\nu^2,
\\
\beta_\nu &\coloneqq -175 + 81\nu + 124\nu^2 + 103\nu^3,
\\
L_r &\coloneqq \log 2 - \frac12 - \log r.
\end{align}
\end{subequations}

\begin{align}
\mathcal{A}_{22}^{\mathrm R}={}&
-\frac{p_r^2}{2}
+\frac{p_\phi^2}{2r^2}
+\frac{1}{2r}
+\frac{p_r^4(5-\nu)}{28}
+\frac{p_\phi^4(-5+\nu)}{28r^4}
+\frac{-4+\nu}{2r^2}
-\frac{547 p_r^3 \nu}{105 r^2}
+\frac{3 p_r^2 (23+\nu)}{28r}
\nonumber\\[0.5ex]
&\quad
+\frac{p_\phi^2(-157+93\nu)}{84r^3}
-p_r\left(
\frac{121 p_\phi^2 \nu}{10r^4}
+\frac{181\nu}{105r^3}
\right)
+\frac{p_\phi^4(962-1375\nu-671\nu^2)}{1008r^5}
+\frac{p_\phi^6(35-13\nu-17\nu^2)}{336r^6}
\nonumber\\[0.5ex]
&\quad
-\frac{p_r^6\alpha_\nu}{336}
+\frac{380-343\nu+205\nu^2}{252r^3}
+\frac{p_\phi^2(-5519-21680\nu+7112\nu^2)}{3024r^4}
+\frac{p_\phi^4(515749+2634215\nu-747870\nu^2-941830\nu^3)}{332640r^6}
\nonumber\\[0.5ex]
&\quad
+\frac{p_\phi^6(-17804+41732\nu+8056\nu^2-3775\nu^3)}{22176r^7}
-\frac{p_r^8\beta_\nu}{2464}
+\frac{p_\phi^8\beta_\nu}{2464r^8}
\nonumber\\[0.5ex]
&\quad
+p_r^2\left(
\frac{p_\phi^2(101-85\nu-313\nu^2)}{168r^3}
+\frac{-1055-2114\nu-52\nu^2}{504r^2}
+\frac{p_\phi^4(35-13\nu-17\nu^2)}{336r^4}
\right)
\nonumber\\[0.5ex]
&\quad
+p_r^4\left(
-\frac{5(85-11\nu+17\nu^2)}{336r}
-\frac{p_\phi^2\alpha_\nu}{336r^2}
\right)
\nonumber\\[0.5ex]
&\quad
+p_r^6\left(
-\frac{p_\phi^2\beta_\nu}{1232r^2}
+\frac{1015-329\nu-376\nu^2+309\nu^3}{1056r}
\right)
\nonumber\\[0.5ex]
&\quad
+p_r^4\left(
\frac{145911-158735\nu+279575\nu^2-17275\nu^3}{55440r^2}
+\frac{p_\phi^2(6425+19793\nu+38624\nu^2+57303\nu^3)}{22176r^3}
\right)
\nonumber\\[0.5ex]
&\quad
+p_r^2\left(
\frac{p_\phi^2(-346391+1446815\nu+6766680\nu^2-2503780\nu^3)}{332640r^4}
-\frac{p_\phi^6\beta_\nu}{1232r^6}
\right.
\nonumber\\
&\qquad\qquad
+\frac{p_\phi^4(-34060+13734\nu+93620\nu^2+56051\nu^3)}{22176r^5}
\nonumber\\
&\qquad\qquad\left.
+\frac{
35662744-22769400\nu+2130975\pi^2\nu-4951200\nu^2-2934200\nu^3+6779520L_r
}{3326400r^3}
\right)
\nonumber\\[0.5ex]
&\quad
+\frac{p_\phi^2}{6652800r^5}
\left(
384920808+240993400\nu-10654875\pi^2\nu-87575200\nu^2+32673200\nu^3+47456640L_r
\right)
\nonumber\\[0.5ex]
&\quad
+\frac{1}{r^4}\left(
\frac{633211}{103950}
-\frac{57481\nu}{2772}
+\frac{205\pi^2\nu}{128}
-\frac{625\nu^2}{462}
+\frac{12071\nu^3}{8316}
+\frac{107}{105}L_r
\right).
\end{align}
\end{widetext}

\begin{widetext}
\begin{align}
\mathcal{A}_{22}^{\mathrm I}={}&
\frac{p_r}{r}
+\frac{p_r^3(-5+\nu)}{14r}
-\frac{838 p_\phi^2\nu}{105r^5}
-\frac{338\nu}{105r^4}
+\frac{92 p_r^2\nu}{35r^3}
-\frac{p_r^5\alpha_\nu}{168r}
+\frac{p_r^7\beta_\nu}{1232r}
\nonumber\\[0.5ex]
&\quad
+p_r\left(
\frac{p_\phi^2(-5+\nu)}{14r^3}
-\frac{185+12\nu}{42r^2}
\right)
\nonumber\\[0.5ex]
&\quad
+p_r^3\left(
-\frac{p_\phi^2\alpha_\nu}{84r^3}
+\frac{695+11\nu+124\nu^2}{252r^2}
\right)
\nonumber\\[0.5ex]
&\quad
+p_r\left(
-\frac{p_\phi^4\alpha_\nu}{168r^5}
-\frac{p_\phi^2(-201-174\nu+32\nu^2)}{168r^4}
-\frac{-1351-2452\nu+523\nu^2}{378r^3}
\right)
\nonumber\\[0.5ex]
&\quad
+p_r^5\left(
\frac{3p_\phi^2\beta_\nu}{1232r^3}
-\frac{25235-4077\nu-5233\nu^2+6164\nu^3}{11088r^2}
\right)
\nonumber\\[0.5ex]
&\quad
+p_r^3\left(
\frac{3p_\phi^4\beta_\nu}{1232r^5}
+\frac{p_\phi^2(-28771-14093\nu-35218\nu^2+7720\nu^3)}{11088r^4}
\right.
\nonumber\\
&\qquad\qquad\left.
+\frac{-368048-150355\nu-179055\nu^2+115625\nu^3}{41580r^3}
\right)
\nonumber\\[0.5ex]
&\quad
+p_r\left(
\frac{p_\phi^6\beta_\nu}{1232r^7}
+\frac{p_\phi^4(-4813-21760\nu+36557\nu^2+14628\nu^3)}{11088r^6}
\right.
\nonumber\\
&\qquad\qquad
-\frac{p_\phi^2(-25967+474545\nu-459810\nu^2+31850\nu^3)}{83160r^5}
\nonumber\\
&\qquad\qquad\left.
+\frac{-261284-176580\nu+7749\pi^2\nu+58700\nu^2-12800\nu^3-30816L_r}{3024r^4}
\right).
\end{align}
\end{widetext}

\section{Second postadiabatic order mismatches for the inspiral only} \label{section:InspiralMismatches}
In this section, we more closely examine the accuracy of the 2PA NIT model by plotting the mismatches for inspiral-only waveforms across parameter space, without handing over to the EOB dynamics to compute the transition to plunge.

\begin{figure*}[tbp]
\includegraphics[width=0.9\textwidth]{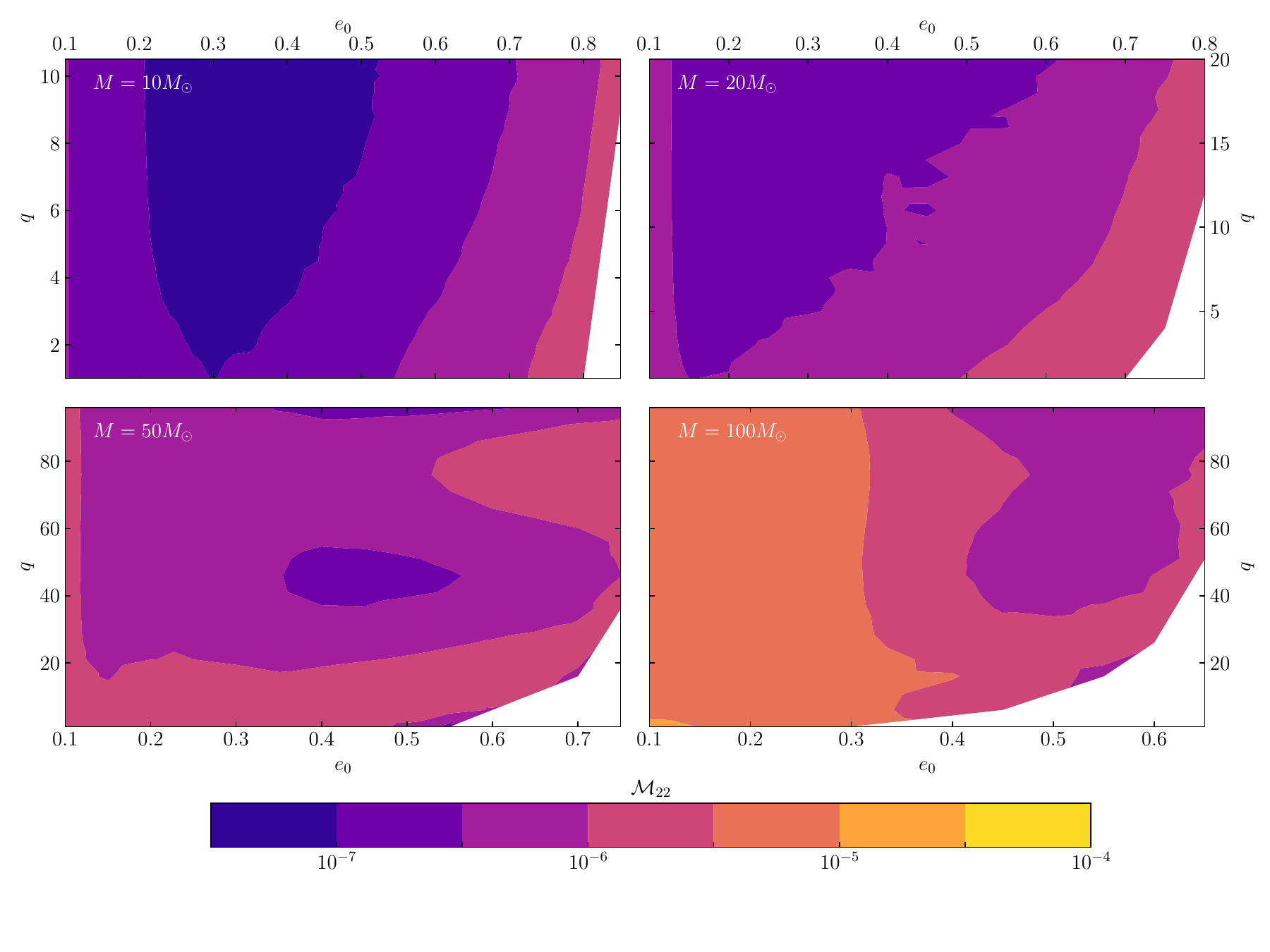}
\caption{Contour plot of the mismatch between EOB and 2PA NIT dynamics for only the inspiral portion of the waveforms for four different total mass values. The mismatch is calculated using the O5 design sensitivity curve and a lower frequency cutoff of 10 Hz. The regions in white correspond to systems that begin after the handover conditions for the NIT dynamics are violated and are excluded from the analysis.
\label{fig:2PA_Mismatch_Inspiral_Contours}
}
\end{figure*}

In Fig.~\ref{fig:2PA_Mismatch_Inspiral_Contours}, we show the second postadiabatic mismatches for a variety of eccentricities and mass ratios for four different total masses, with the same setup as in Fig.~\ref{fig:2PA_Mismatch_Contours} but with the waveforms truncated when one of the three mismatch criteria is met.

This removes the errors that might occur when transforming from the NIT variables to the EOB variables when computing the transition to plunge. We see that the mismatches now vary smoothly across the parameter space, and are generally lower than in Fig.~\ref{fig:2PA_Mismatch_Contours} with $\mathcal{M}_{22} \lesssim 10^{-5}$. For all masses, we see slightly higher mismatches for the lowest eccentricities than at moderate eccentricities. We believe this is due to the poor convergence of the parametrization as one approaches the quasicircular limit, especially for the most relativistic systems with $M = 100M_\odot$. 

Outside of this, we see that the mismatch generally follows the expected trend of getting worse for higher eccentricities and smaller mass ratios, corresponding to systems with lower adiabaticity.

\bibliography{EOBNITsEccentricReferences.bib}

\end{document}